\documentclass{emulateapj}
\usepackage{times}

\lefthead{OKOSHI et al.}
\righthead{HI-selected Galaxies as a probe
of Qusar Absorption Systems}

\begin{document}
\title{HI-selected Galaxies as a probe
of Quasar Absorption Systems}

\author{Katsuya Okoshi}
\affil{
Tokyo University of Science,
Oshamanbe, Hokkaido, 049-3514,Japan: okoshi@rs.kagu.tus.ac.jp}
\author{Masahiro Nagashima}
\affil{
Nagasaki University,
Nagasaki, 852-8521, Japan}
\author{Naoteru Gouda}
\affil{National Astronomical Observatory of Japan,
Mitaka, Tokyo, 181-8588,  Japan}
\and
\author{Yousuke Minowa}
\affil{Subaru Telescope,
National Astronomical Observatory of Japan, 
650 North A'ohoku Place 
Hilo, HI 96720, USA }

\begin{abstract}

We investigate the properties of HI-rich galaxies detected in blind radio 
surveys within the hierarchical structure formation scenario using a semi-analytic 
model of galaxy formation.  
By drawing a detailed comparison between the properties of HI-selected 
galaxies and HI absorption systems, we argue a link between the local 
galaxy population and quasar absorption systems, particularly for damped 
Ly$\alpha$ absorption (DLA) systems and sub-DLA systems. 
First, we evaluate how many HI-selected galaxies exhibit HI column densities 
as high as those of DLA systems. 
 We find that HI-selected galaxies 
with HI masses $M_{\rm HI} \ga 10^{8} M_{\odot}$ have gaseous disks 
that produce HI column densities comparable to those of DLA systems.  
We conclude that DLA galaxies where the HI column densities 
are as high as those of DLA systems,  contribute significantly to the population 
of HI-selected galaxies  at $M_{\rm HI} \ga 10^{8} M_{\odot}$. 
Second, we find that star formation rates (SFRs) correlate tightly with 
HI masses ($M_{\rm HI}$) rather than $B$- (and $J$-) band luminosities: 
$SFR \propto M_{\rm HI}^{\alpha}$,  $\alpha=1.25-1.40$ for  
$10^{6} \le M_{\rm HI}/M_{\odot} \le 10^{11}$ . 

In the low-mass range $M_{\rm HI} \la 10^{8} M_{\odot}$, sub-DLA galaxies  
replace DLA galaxies as the dominant population. 
The number fraction of sub-DLA galaxies relative to galaxies reaches  
$40\% - 60$ $\%$ for $M_{\rm HI} \sim 10^{8} M_{\odot}$ and $30\% 
 -80 \%$ 
for $M_{\rm HI} \sim 10^{7} M_{\odot}$.  
The HI-selected galaxies at $M_{\rm HI} \sim 10^{7} M_{\odot}$ are a strong probe 
of sub-DLA systems that place stringent constraints on galaxy formation and evolution.

\end{abstract}



\section{Introduction}

The absorption-line systems found in quasar/gamma-ray burst spectra, the 
so-called "quasar absorption systems",  provide us with a unique probe of galaxy 
formation and evolution processes. The systems offer valuable opportunities to  
explore the physical and chemical properties of inter-galactic media and/or 
intervening galaxies (e.g., the amount of neutral gas and metals, the gas dynamics).  
In particular, because the absorption feature provides 
basic information about hydrogen gas which is a main component of galactic gas, 
HI absorption systems have been studied extensively as means of placing  
stringent constraints on galaxy formation and evolution.  

Recently, radio surveys have provided large samples of HI-selected galaxies and 
significantly improved our understanding of galaxy evolution. 
The HI-selected galaxies have been explored 
in blind surveys of the HI emissions, for example, the Slice Survey (SS; e.g., 
Spitzak \& Schneider 1998), the Arecibo Dual-Beam Survey(ADBS; e.g., Rosenberg \& Schneider 
2000), the HI Parkes All Sky Survey (HIPASS; e.g., Zwaan et al. 2003; Meyer et al. 2004), 
the HIDEEP survey (HIDEEP; e.g., Minchin et al. 2003), the HI Nearby Galaxy Survey 
(THINGS; Walter et al. 2008),  and the Arecibo Legacy Fast ALFA 
survey (ALFALFA; e.g., Giovanelli et al. 2005). 
These surveys result in an accurate measurement of the basic properties of galactic HI-gas
(e.g., the HI mass function,  HIMF).  When the HI-selected galaxies exist on a line of sight 
to quasar/gamma-ray burst, the HI-selected galaxies can be 
 HI-absorption systems due to having the HI column densities 
as high as those of  HI absorption systems. In this paper, using a model of  galaxy 
formation, we aim to reveal the basic properties of HI 
absorption systems by focusing on HI-selected galaxies in blind radio surveys. 

Among the HI absorption systems, 
damped Lyman $\alpha$ absorption (DLA) systems have been studied because 
the high HI column densities ($N_{\rm HI} \geq 10^{20.3}$ cm$^{-2}$) are 
considered to be 
a good probe of an early stage of star formation processes.  The origin of DLA systems 
has been argued on the basis of the observational properties.  It has been considered as, 
for example,  rapidly rotating gaseous disks of massive 
spiral galaxies (e.g., Wolfe et al. 1986), gas-rich dwarf galaxies (e.g., York et al. 1986), 
and so on. 

Theoretical studies, utilizing numerical simulations and semi-analytic models, have been 
tackled to revealing the nature of DLA systems and a relationship to galaxy population 
 \citep[e.g.][]{Katz96, Kau96, H98, G01, MPSP01, ONGY, ON05, N04, N07, JE06, 
Raz06, Raz08,  Po08, BH09, T09}. 
 Although the studies reproduce many properties of DLA systems, definite conclusion has not   
 been reached yet. For example, 
 at high redshifts ($z>2$), DLA systems have high number densities per unit redshift 
 (the incidence rate) $dN/dz$ and large velocity widths $\mathit{\Delta} V$ of the absorption lines. 
 Using models under an assumption that DLA systems arise mainly in galactic disks of 
virialized halos, 
 the high number densities require large number of less massive galaxies. However, 
 the predicted velocity widths $\mathit{\Delta} V$ of absorption lines produced by 
less massive systems appear to be smaller than the  typical values of observed ones \citep{PW97}.  
This suggests that the high number densities are also attributed to large cross sections 
of the other population,   
for example, extended gas that resides within/outside virial radii of halos, outflow gas 
due to supernova explosions, tidal gas related to galaxy interactions, and so on. 
 At high redshifts, the mixed population, in addition to compact disks, may give rise to 
 damped Lyman $\alpha$ absorption lines.  
 
 There are the dedicated attempts to detect the host galaxies of DLA systems by optically 
deep-imaging observations (e.g., Steidel et al. 1994, 1997; Le Brun 
et al. 1997; Rao \& Turnshek 1998, 2000; Turnshek et al. 2001; Bouch$\rm \acute{e}$ et al. 2001; 
Bowen, Tripp \& Jenkins 2001; Warren et al. 2001; M$\rm \o$ller et al. 2002, 2004; Fynbo 
et al. 2003; Rao et al. 2003; Chen \& Lanzetta 2003; Lacy et al. 2003; Schulte-Ladbeck et 
al. 2004; Chun et al. 2006; Gharanfoli et al. 2007).  
A few cases of successful identifications suggest that the host galaxies of DLA 
systems present mixed morphological types of galaxies. 
However, in many cases, it is a challenging task to identify the host galaxies in the emission  
of the optical surveys, because  many host galaxies are very faint and/or very compact. 

At low redshifts ($z<1$), it is expected that more DLA hosts can be identified even in the 
case when their luminosities are low. In the observation,  
for example, \citet{R03} have compiled data for host galaxies at low 
redshifts ($z<1$).  However,  their sample size is still small ($\sim 14$).  
They argued that the sample is dominated by galaxies with low-luminosities and small impact 
parameters. 
\citet{ON05} have theoretically investigated galaxies 
having HI column densities as high as those of DLA systems (hereafter, 
DLA galaxies) at low redshifts using a semi-analytic galaxy formation model.  
We concluded that DLA galaxies consist primarily of low luminosity galaxies with small 
sizes (typical radius $\sim 3$ $h^{-1}$kpc,
surface brightness in the $B$-band $22$ to $27$ mag arcsec$^{-2}$). 
This result is consistent with the observations that, at low redshifts, DLA systems arise 
primarily in galactic disks rather than tidal gas caused by galaxy interactions or gas 
driven by galactic winds. 
Moreover, in our conclusion, 
(1) DLA galaxies have typical star formation rates (SFRs) 
$\sim 10^{-2}$ $M_{\odot}$ yr$^{-1}$  
and (2) that the difficulty in identifying DLA galaxies is due not only to 
their faintness but also to a {\it masking effect} which occurs when the point 
spread function (PSF) of the respective background quasars masks the DLA galaxies. 
We suggest that 
$60\% -90 \%$ of DLA galaxies suffer from this masking effect. 
Thus,  DLA galaxies in the local universe significantly should consist of the optically-faint and 
compact systems that require another approach to host galaxy detection in observations.  

To understand DLA systems, 
it is useful to investigate HI-selected galaxies detected 
in radio surveys because they contain a large amount of neutral gas that potentially gives rise to 
strong absorption lines when bright emission sources exist behind them. 
The ADBS, HIPASS, and THINGS provide significantly large samples  for DLA galaxies 
compared to those of the optical counterparts of DLA systems 
identified in the optical and near-infrared images at $0 \le z \le 1$ \citep[e.g.][]{R03}. 
The large radio samples show  interesting properties (e.g., HI cross-section and HI mass ) of 
DLA galaxies   \citep[e.g.][]{Rao93, RS03, Ryan03, Z05b, Z08}.   
Thus, the study of radio samples can offer {\it statistically} valuable information 
about DLA galaxies and opportunities for exploring the evolution of DLA systems.   
The advantage of using the radio samples is summarized as follows: 
(1) the radio survey has no bias against  optically low surface brightness galaxies  
that may be missed by optical and near-infrared surveys, 
(2) the identification process is free from the masking effect because 
no bright sources are located behind them, and (3) the sample size is much larger 
than those of DLA optical counterparts. 

By employing our model for DLA galaxies, we investigate the number of HI-selected galaxies 
that can be DLA galaxies.  Next, we focus on the optical properties  (e.g., SFR ) of DLA galaxies 
to explore how DLA galaxies should be detected in optical surveys.   Finally, we investigate   
the number of HI-selected galaxies that can be host galaxies of other 
quasar absorption systems such as sub-DLA systems that exhibit HI column 
densities lower than those of DLA systems. This study of HI-selected galaxies offers 
valuable opportunities for exploring the low-redshift end of the evolution of DLA and 
sub-DLA systems.  \\

In Section 2, we describe our models used in this paper. In Section 3, we argue a relationship  
between the HI-selected galaxies and DLA galaxies.  
In Section 4, we investigate the optical/infrared properties of the HI-selected galaxies 
and DLA galaxies. In Section 5, we focus on another 
population of absorption systems, namely sub-DLA systems, and argue how they 
contribute to the local galaxy population. In Section 6, we derive several 
implications from our results. Finally, we draw our conclusion in Section 7.


\section{Model}

We employ a semi analytic model of galaxy formation for DLA galaxies 
\citep[][hereafter Paper I]{ONGY}. It has been shown that  
this model accounts for the observational properties of local galaxies, such as luminosity 
functions and cold gas mass fractions \citep{NTGY}.  Previously, we have shown that  
our model of DLA galaxies reproduces several optical/near-IR observations for the host 
galaxies of DLA systems at low redshift $z < 1$ \citep[][hereafter Paper II]{ON05}. 
We have also found a tight correlation between cross sections  and 
HI masses of DLA galaxies entirely consistent 
with the ADBS result \citep{RS03}. In this paper, we adopt almost the same model as  
in Paper II and investigate DLA galaxies at redshift $z=0$ by comparing 
radio and optical properties between DLA galaxies and HI-selected galaxies.  
The outline of this model is described as follows.

We adopt a standard $\Lambda$CDM cosmological model with parameters, 
$\Omega_{0}=0.3$, $\Omega_{\Lambda}=0.7$, $\Omega_{\rm b}=0.015h^{-2}$, 
$h=0.7$ (where $h \equiv H_{0}/100$ km s$^{-1}$ is the Hubble parameter), and 
$\sigma_{8}=1$, the normalization of the power spectrum of density fluctuations given 
by \citet{BBKS}.  The number of progenitors
of a dark halo is given by an extended Press-Schechter model. The mass function at 
$z=0$ is provided by the Press-Schechter mass function \citep{PS74, BCEK, B91, 
LC93}. The realizations of progenitor histories are computed by the method developed 
by \citet{SK99}. We focus on halos with circular velocity $V_{\rm circ} \ge 30$ km s$^{-1}$ 
because we assume that less massive systems ($V_{\rm circ} \le 30$ km s$^{-1}$) correspond 
to diffuse accretion masses. The main difference from the previous model is a value adopted  
for the threshold on the accretion mass, $V_{\rm circ,th}$; $V_{\rm circ,th}$ $= 40$ km 
s$^{-1}$ in Papers I and II. 

We assume that baryonic gas comprises cold and hot phases. When a halo collapses, halo 
gas is assumed to be shock-heated to the virial temperature of the halo and distributed 
in a singular isothermal sphere ({\it hot gas}).  The gas component partly cools by 
radiative cooling processes. The cooled gas is called `{\it cold gas}'. Here we use a 
metallicity-dependent cooling function (Sutherland \& Dopita 1993). After the cold gas 
falls onto a central region of the halo, stars are formed from the cold gas. 
The SFR  is assumed to be $\dot{M}_{*}=M_{\rm cold}/\tau_{*}$,  
where $M_{*}$ and $M_{\rm cold}$ are the masses of stars and cold gas, respectively, 
and $\tau_{*}$ is the timescale of star formation.  We simply assume that the SFRs are 
constant in the galactic disks. Following our previous models, we assume the timescale 
of star formation; $\tau_{*} = \tau_{*}^{0}(V_{\rm circ}/V_{*})^{\alpha_{*}}$. 
In Papers I and II, we found that, to account for several observational properties of the DLA hosts, 
$\tau_{*}$ should be constant with redshift and dependent on the circular velocity 
$V_{\rm circ}$ of the halos hosting DLAs,  i.e., ($\alpha_{*}$, $V_{*}$)=($-2$, $300$ 
km s$^{-1}$). Here we determine the parameters ($\alpha_{*}$, $V_{*}$) by matching 
the HI mass function to those that have been observed. Initially, we adopt a set of parameters 
($\alpha_{*}$, $V_{*}$)=($-2$, $300$ km s$^{-1}$) 
and $\tau_{*}^{0}=1.5$ Gyr, the same set as used for the previous model 
(Constant Star formation model) in Papers I and II. 
 
Moreover, we also consider another model for star formation, 
adopting a different set of parameters ($\alpha_{*}$, $V_{*}$). Assuming that the star 
formation timescale $\tau_{*} \propto V_{\rm circ}^{\alpha_{*}}$, adopting small  
(large negative)  values of $\alpha_{*}$  indicates that the star formation timescales are 
longer in less massive 
galaxies. This leads less massive galaxies to have lower SFRs and then  
produces high number densities of low $M_{\rm HI}$ galaxies. Here we employ a 
model in which the star formation rates are low in less massive systems by adopting 
$\alpha_{*}=-3$.  Hereafter, we call the two models:  High Star formation (HS) model 
with $\alpha_{*}=-2$ and Low Star formation (LS) model with $\alpha_{*}=-3$. 
The normalizations of the star formation timescale, 
(${\tau_{*}}^{0}$, $V_{*}$)=($1.5$ Gyr, $300$ km s$^{-1}$), are 
identical for both models. 

Supernova feedback processes are incorporated in this model. The rate of reheating 
by supernova feedback is assumed to be $\dot{M}_{\rm reheat}=\beta \dot{M_{*}}$. 
Because the feedback process should affect the physical condition of 
the gas more drastically in less massive galaxies, we assume that 
$\beta=(V_{\rm circ}/V_{\rm hot})^{-\alpha_{\rm hot}}$, where the parameter sets, 
($V_{\rm hot}$, $\alpha_{\rm hot}$)=($240$ km s$^{-1}$, $2.5$)
\footnote{
We here adopt a parameter $V_{\rm hot} = 240$ km s$^{-1}$ slightly smaller than 
$280$ km s$^{-1}$ adopted in Papers I and II because this better matches the 
observational data of HI mass functions as presented below. We have confirmed that 
this makes little difference to our previous results in Papers I and II.  
}
 for the HS model 
and ($320$ km s$^{-1}$, $2.5$) for the LS model, are required to 
account for the observed optical luminosity functions of local galaxies. 
The results for optical luminosity functions are argued in Section 3.

We include the merging of galaxies in this model. In our model, when two or more 
dark halos merge, the central galaxy in the largest progenitor halo becomes the new 
central one. The other galaxies become satellites dwelling in the largest halo of the 
central galaxy. 
We assume that the merging of  {\it galaxies} in a common halo occurs due to two types 
of mechanisms: dynamical friction and random collision. When satellite galaxies merge with 
the central one, the merging occurs on the dynamical friction timescale.
In the merging of satellites, they collide randomly within the mean-free timescale. 
After the merging of two galaxies (e.g., A and B), a starburst occurs and all 
cold gas is consumed by the star formation if the mass ratio, $m_{\rm A}/m_{\rm B}$ 
($m_{\rm B}>m_{\rm A}$), is larger than 
$f_{\rm bulge}$ ({\it major merger}). Otherwise, no star formation activity occurs when 
the mass ratio is smaller than $f_{\rm bulge}$ ({\it minor merger}).  
In this paper, we adopt $f_{\rm bulge}=0.5$ in the same way as in Papers I 
and II.

We address the definition of the size and the HI column density of a DLA galaxy. 
In this model, the radial distribution of the HI column density follows an exponential
profile with an effective radius of a gaseous disk, $r_{\rm e}$, obtained by assuming 
the specific 
angular momentum conservation of the cooling hot gas. The dimensionless spin parameter 
of dark halos has a log-normal distribution with an average of $0.06$ and a logarithmic 
variance of $0.6$. Using the above assumptions, we calculate the following quantities.  
The first is    
the central column density provided by $N_{\rm 0}=M_{\rm cold}/(2 \pi \mu m_{\rm H} 
r_{\rm e}^{2})$, where $m_{\rm H}$ is the mass of a hydrogen atom and $\mu(=1.3)$ is the 
mean molecular weight. In this procedure, in order to take into account the inclination, 
$\mu_{\rm inc} = \cos \theta$ of gaseous disks in each galaxy, we create random 
realizations of the inclinations, and then calculate the cross section and the column density.  
The second is the size of a DLA galaxy defined by a radius $R$ where  
$N_{\rm HI} = 2 \times 10^{20}$ cm$^{-2}$.  Here we adopt the conventional 
threshold for the HI column density of DLA systems.  
The third is the HI column density of a DLA galaxy, which is defined as the column density 
averaged over radius within $R$.

\section{HI-selected Galaxies and DLA Galaxies}

First, we present the optical luminosity function and 
the HI mass function (HIMF) of local galaxies provided by the optical and 
the radio surveys. 
These properties are strong indicators used to probe the physical conditions of 
nearby galaxies with statistically significant confidence because, 
for the optical luminosity function, the sample size of the Sloan Digital Sky 
Survey  (SDSS)   
is quite large,  more than 30,000  (Blanton et al. 2005), and,  for the HIMF, 
the sample size from the HIPASS is about $4000$  (Zwaan et al. 2005a) . 

\subsection{Luminosity Function}

In Figure 1, we show the luminosity function  at redshift 
$z=0$.  Figure 1(a) shows the $B$-band luminosity function of galaxies 
 (solid line) in the $r$ band for the HS model. The observational data are given 
by the SDSS  Data Release 2  
(filled circles; Blanton et al. 2005). 
We set the  supernova feedback-related parameters, $\alpha_{\rm hot}$ and 
$V_{\rm hot}$, so as  
to account for the faint-end slope of luminosity function that is sensitive to 
the supernova feedback because it predominantly suppresses the 
formation of dwarf and faint galaxies. 
It appears 
that the abundance of bright galaxies ($M_{\rm r} \la -20$) is slightly higher  
than the observation. This is partly because, for simplicity,  we adopt total magnitudes 
instead of the Petrosian magnitudes that the SDSS team uses. This makes simulated galaxies 
more luminous than the observed ones.   

In Figure 1(b), we show the $r$-band luminosity function of galaxies 
 (solid~line) for the LS model.  
The LS model reproduces the faint-end slope of the luminosity 
function. However, the abundance of bright galaxies, particularly at 
$M_{\rm r} \sim -20$, is slightly lower than that for the HS model that reproduces 
the observation.  
For the LS model, by adopting $V_{\rm hot}= 320$ km s$^{-1}$ instead of 
$240$ km s$^{-1}$ used for the HS model, the reheating rate by supernova feedback is 
higher than that for the HS model.  
The supernova feedback suppresses the formation of bright galaxies more drastically  
than the HS model.

In Figures 1( a) and 1(b), we also show the optical luminosity functions of {\it 
DLA galaxies} (dashed lines) for the HS and LS models, respectively. 
Here, `DLA galaxies' are selected by a criterion where 
the gaseous disks exhibit HI column densities above the threshold, 
$N_{\rm HI} = 10^{20.3}$ cm$^{-2}$, while `galaxies' are defined without any thresholds 
on the HI column density.  The shapes of luminosity 
functions of DLA galaxies are similar to those of galaxies in the whole range of the observed 
magnitudes, indicating that many (bright and faint) galaxies have gaseous 
disks with HI column densities large enough to give rise to damped Ly$\alpha$ 
absorption lines when they are on a line of sight to the background 
quasar. In other words, the low-redshift DLA systems arise not only in bright 
$L^{*}$ galaxies but also in faint ones.

\subsection{HI mass function}

The HI mass function (HIMF) is one of the most important statistical properties of local 
galaxies. While the optical luminosity function describes the distribution of stellar mass, 
the HIMF describes the distribution of cold gas mass. The HIMF has an advantage 
in terms of  the ability to trace the HI-mass distribution of galaxies even in the 
case where the optical luminosities are too faint to identify  in optical surveys. 

Figure 2(a) shows the HIMF of galaxies for the HS model (solid line). 
The predicted HIMF 
is similar to a Schechter-type function. Blind 21cm surveys have shown that 
HI-selected galaxies follow an HIMF, $\theta$, that can be well fitted by a Schechter 
function with a faint-end slope $\alpha$, $\theta \propto M_{\rm HI}^{\alpha}$, 
e.g., for the HIPASS sample, $\alpha=-1.37$ in the mass range 
$10^{7.2} \le M_{\rm HI}/M_{\odot} \le 10^{10.7}$ \citep{Z05a}. However, because the
survey fields and/or the sample sizes are different, the slopes of HIMFs still 
appear to differ from survey to survey (e.g., $\alpha=-1.1$ to $-1.6$).  
This may be due to environmental effects in which the slope depends on the 
local galaxy density. Taking into account this 
uncertainty, we present the range of HIMFs provided by various blind surveys 
(the shaded region;  Henning et al. 2000; Springob, Haynes \& Giovanelli 2005; 
Rosenberg \& Schneider
2002, 2003; Zwaan et al. 1997, 2003, 2005a), together with 
some individual data for the HIMF (open squares, Rosenberg \& Schneider 2002; filled circles, Zwaan et al. 2005a). 

The HS model produces an abundance of galaxies within the range of the observational 
data.  Particularly at $M_{\rm HI} \sim 10^{9}-10^{9.5}$ $M_{\odot}$
where the sample size provided by the HIPASS is the largest in the number density  
(Zwaan et al. 2005a, 2005b), our result is consistent with the observations.  
However, in the low-mass end,  
the predicted abundance is somewhat lower than the HIPASS observations. 
Although the small observational 
sample at  $M_{\rm HI} < 10^{8} M_{\odot}$ accounts for some uncertainties in the 
low-mass end, for the HS model, the underprediction is mainly due to the  
high star formation rates suppressing the formation of low $M_{\rm HI}$ 
galaxies.  

In Figure 2(b), we show the HIMF for the LS model (solid~line).  
The LS model also produces a good match to the observations.  
In comparison to the data range of number densities $\theta$  
at $M_{\rm HI} \sim 10^{9}$ $M_{\odot}$, the LS model produces 
an abundance of galaxies corresponding to upper 
limit of observed number densities, while the HS model produces one corresponding to lower limit.

In Figure 2, we also show the HIMFs of {\it DLA galaxies} 
(dashed~lines).  For the HS and the LS models, the HIMFs of DLA galaxies 
are obviously similar to those of galaxies 
(in Figure 2(b), the HIMF is almost identical to those of galaxies). This result 
suggests that,  in the radio blind surveys, 
the HI-selected galaxies  mostly have 
gaseous disks with the HI column densities as high as those of DLA systems. 
However, it should be noted that for the HS model, we find that the number densities 
of DLA galaxies are somewhat smaller than those of galaxies at the low-mass end. 
This shows that all galaxies do not 
have gaseous disks with HI column densities that are above the DLA threshold.

\subsection{Cold gas mass to $B$-band Luminosity}

In Figure 3, we show the ratios of cold gas mass relative to the $B$-band 
luminosity of spiral galaxies for the HS (solid line) and the LS (dashed 
line) models, respectively. Symbols indicate the observations provided by  
Huchtmeier \& Richter (1988; filled circles), Sage (1993; open circles), 
and Garnett (2002; open squares). The HI-selected galaxies from the HIPASS 
are also included; Koribalski et al. (2004; shaded region) and Warren, Jerjen \& 
Koribalski (2006; filled triangles). The HIPASS data from Koribalski et al. 
(2004) represent  {\it the shaded region} because the sample size is large. The cold gas masses given by the HIPASS and Huchtmeier \& Richter (1988) include only HI.       

The results for the HS and LS models show that faint galaxies have large mass fractions of 
cold gas. The mass fractions of faint galaxies for the LS model are larger than 
those for the HS model 
because the star formation timescales are longer in faint galaxies. Our result shows 
that the predicted cold gas mass fractions are within the range of observational 
data. However, taking into account the addition of the HIPASS data,  
the cold gas mass fraction was still unable to constrain the star formation 
timescale because the range of observed data spans one or two orders of magnitude,
particularly at the faint end. When the samples of faint galaxies increase in the 
future, the ratio $M_{\rm HI}/L_{\rm B}$ of faint galaxies ($M_{\rm B} \ga -16 $) 
will constrain the star formation rates more stringently.   

\subsection{Number Fraction}

For the purpose of evaluating more precisely how many DLA galaxies contribute to 
the galaxy population, in Figure 4, we present number fractions of DLA galaxies   
as a function of HI mass $M_{\rm HI}$. The number fraction at each bin of HI mass 
is defined as a ratio  of the number densities of DLA galaxies relative to those of galaxies 
per bin of HI mass. The result for the HS model (solid line)  shows that the number 
fractions are almost unity at $M_{\rm HI} \ga 10^{8} M_{\odot}$ , which suggests 
a trend that the gaseous disks of massive galaxies do have high enough HI column densities 
to give rise to damped Ly$\alpha$ absorption lines if a background bright 
emission source exists along a line of sight. Noticeably, in the low-mass range 
$M_{\rm HI} \la 10^{8} M_{\odot}$, the number fractions begin to decrease toward 
the low-mass end, e.g., $\sim 0.7$ at $10^{7.5} M_{\odot}$ and $\sim 0.6$ at $10^{7}
M_{\odot}$. We find that, {\it at $M_{\rm HI} \la 10^{8} M_{\odot}$, the number 
distribution of DLA galaxies is not identical to that of galaxies}. This indicates 
that the small amount of HI gas contained in the less massive systems does not 
have high enough HI column densities to detect as DLA systems.
For the HS model, at the low-mass end, another population becomes dominant  
instead of DLA galaxies. The population at the low-mass end is argued in 
Section 5.  

The number fractions for the LS model are also shown in Figure 4 (dashed line). 
While the number fractions are almost unity at $M_{\rm HI} \ga 10^{8} M_{\odot}$, 
they begin to decrease gradually toward the low-mass end ($M_{\rm HI} < 10^{8} M_{\odot}$). 
In comparison to the result for the HS model, the SFRs of less massive galaxies 
are low for the LS model. The low SFRs make the HI column densities of 
low-$M_{\rm HI}$ galaxies as high as those of DLA systems. This increase 
in HI column density causes the number fractions of DLA galaxies to be large   
even at the low-mass end $M_{\rm HI} < 10^{8} M_{\odot}$. 
The differences between the 
numbers of DLA galaxies and galaxies are not as large as those for the HS model. 
However, for the LS model, we find a trend similar to one for the HS model, 
in which not all galaxies correspond to DLA galaxies at the low-mass end 
(e.g., number fraction $\sim 0.9$ at $M_{\rm HI} \sim 10^{6} M_{\odot}$).

\section{Optical Properties of HI-selected Galaxies}

Here, we investigate the difference between galaxies and DLA galaxies through 
the optical/infrared properties of HI-selected galaxies. 
It should be noted that galaxies  
in our calculation are volume-limited, whereas the observed HI-selected galaxies 
are not volume-limited but flux-limited . 
To draw a detailed comparison between them,  attention needs to be paid to the difference 
in the selection criteria. This is an issue that deserves further investigation.  

In Figure 5, we show the predicted distribution of galaxies and DLA galaxies 
in the plane of the HI mass and the luminosity by contours 
for the HS model. In Figure 5(a), we present the $J$-band 
luminosities of the galaxies by contours, together with the observational 
data (filled circles, Rosenberg \& Schneider 2000;  open circles,  
Spitzak \& Schneider 1998). Our result 
is broadly consistent with the observations. In the blind surveys, it has 
been investigated whether the HI-selected galaxies 
exhibit a correlation between the HI mass and the $J$-band luminosity  
\citep{RS03, RS05}. Although our result indicates that  
 at $M_{\rm HI} \ga 10^{8}$ $M_{\odot}$   
there is a correlation between the luminosities in the $J$ band and the HI masses, 
there is no linear relationship between $\log$ $L_{\rm J}$ and $\log$ $M_{\rm HI}$.

The abundance of less massive galaxies ($M_{\rm HI}$ $<$ $10^8$ $M_{\odot}$) 
in the $L_{\rm J}$-$M_{\rm HI}$ plane severely depends on formation processes of 
less massive galaxies.  In our model, we treat halos 
with $V_{\rm circ} < V_{\rm circ,th}=30$ km s$^{-1}$ as a diffuse accretion mass. 
We assume that a suppression of galaxy formation within halos with the circular 
velocities less than $V_{\rm circ,th}$ occurs mainly due to the ionizing 
background radiation. The presence of the ionizing radiation prevents the gas from 
collapsing into a dark halo because the radiation raises the temperature and pressure 
of the gas. As a result of previous studies using hydrodynamic simulations, 
the suppression mechanism should be large in low-mass halos with 
$V_{\rm circ} \la 30-60$ km s$^{-1}$ (e.g., Thoul \& Weinberg 1996; 
Quinn, Katz \& Efstathiou 1996; Weinberg, 
Hernquist \& Katz 1997; Navarro \& Steinmetz 1997; Gnedin 2000; Bullock, Kravtsov \& 
Weinberg 2000; Kitayama \& Ikeuchi 2000; Hoeft et al. 2006; Okamoto, Gao, \& Theuns 2008). 
The suppression of galaxy formation in the 
low-mass halos with $V_{\rm circ} < V_{\rm circ,th}=30$ km s$^{-1}$ reduces 
the number densities of less massive galaxies. This directly affects the shape of 
the luminosity function of galaxies by making a shallower faint-end slope
\citep{N99,S02,B02}. In the $L_{\rm J}$-$M_{\rm HI}$ plane, the luminosity 
distribution, particularly at the faint end, 
also depends sensitively on $V_{\rm circ,th}$. If the suppression threshold,  
$V_{\rm circ, th} < 30$ km s$^{-1}$,  the abundance of faint galaxies 
( $L_{\rm J}  \la 10^{8} L_{\odot}$) shows a significant increase. 
The abundance of less massive galaxies is a potentially useful probe that places 
constraints on $V_{\rm circ,th}$.  

In Figure 5(b), we show the number distributions of DLA galaxies in the 
$L_{\rm J}$-$M_{\rm HI}$ plane by contour lines for the HS model. The $J$-band 
luminosities are broad and span approximately four orders of magnitude, $10^{7}$ 
to $10^{11}$ $L_{\odot}$ at $M_{\rm HI}$ $\sim 10^{9}$ $M_{\odot}$.  
The luminosity distribution of DLA galaxies does not significantly differ from that of 
galaxies (Figure 5(a)). We find that, for each bin of HI mass 
($M_{\rm HI}$ $\ga $ $10^{8}M_{\odot}$), 
the averaged luminosities $L_{\rm J}$ of galaxies match those of DLA galaxies 
within one $\sigma$ of error. The main reason is that most galaxies have gaseous 
disks where the HI column densities above the DLA threshold, 
particularly at $M_{\rm HI} \ga 10^{8} M_{\odot}$ (see Figure 4).

In Figure 6, for the LS model, we plot the abundances of galaxies 
(Figure 6(a)) and DLA galaxies (Figure 6(b)) in the plane of the HI mass and 
the $J$-band luminosity.  
In Figure 6(a), our result is consistent with the observations, particularly 
at $M_{\rm HI} \ga 10^{9} M_{\odot}$. It is evident that the number distributions 
in the $M_{\rm HI}$-$L_{\rm J}$ planes show no difference between DLA galaxies and 
galaxies because most galaxies correspond to DLA galaxies for the LS model as shown in 
Figure 4. 
Both models successfully reproduce the observed abundance of galaxies at the bright end of 
$J$-band luminosities ($L_{\rm J} \sim 10^{10} L_{\odot} $). 
However, our models produce wide spreads at the faint end 
($L_{\rm J} \la 10^{8} L_{\odot} $) in the $M_{\rm HI}$-$L_{\rm J}$ planes. 
Thus, the wide spreads cannot put stringent constraints on the star formation rates.

Figures 5(c) and 6(c) show the $B$-band luminosity distributions of galaxies 
as a function of HI mass for the HS and LS models, respectively. Obviously, the 
$B$-band luminosity correlates with the HI mass more tightly than the $J$-band 
luminosity because the $B$-band luminosity is more acutely sensitive to the SFR  than 
the $J$-band luminosity. In observations, this correlation has been suggested by a survey 
for optical counterparts of the HI-selected samples. For example,  \citet{D05} found 
the optical counterparts of $\sim 84 \%$ among   
the sample of $\sim 4300$ HI-selected galaxies. They argued that 
there is a clear trend that more luminous galaxies have higher radio fluxes.  This is 
consistent with our result.  The results can be of great interest in comparing the radio properties 
of nearby galaxies to the optical (e.g., the $B$-band ) ones.

Furthermore, we focus on a relationship between SFRs and HI masses in galaxies. 
In our model, the SFR is proportional to the cold mass with a typical 
timescale $\tau_{*}$ as $\dot{M}_{*}=M_{\rm cold}/\tau_{*}$. This means that the 
star formation timescale is induced directly from the slope of the 
SFR-$M_{\rm HI}$ relation. Figure 7 
shows the SFRs in galaxies as a function of HI mass. Figure 7(a) presents the 
averages of SFRs in galaxies with one $\sigma$ error bars for the HS model (red 
solid-line) and the LS model (blue solid-line), respectively. 
It is evident that massive galaxies exhibit high SFRs. When this relation SFR 
$\propto$ $M_{\rm HI}^{\alpha}$ is fitted by averaged least squares, we find the 
slope $\alpha=1.25 \pm 0.05$ (the HS model) and $1.40 \pm 0.07$ (the LS model) in the range 
of HI mass $10^{6} \le M_{\rm HI}/M_{\odot} \le 10^{11}$. Moreover, we also show 
the SFRs in DLA galaxies as a function of HI mass for the HS model (red  dashed-line) 
and the LS model (blue dashed-line), respectively. The average SFRs in DLA galaxies 
are almost identical to those in galaxies. The LS model predicts 
the SFRs lower than the HS model because, for the LS model, the star formation timescales 
are longer particularly for less massive galaxies. 

Surveys for emission lines from nearby galaxies have been attempted to investigate 
the star formation activity. For example, the emission-line images in the H$\alpha$ band 
provide valuable measurements of the ionizing flux against the interstellar medium (ISM) 
and of the SFRs in the ISM (e.g., Kennicutt 1983, 1998, and references therein). Because large samples of HI-selected 
galaxies obtained from the blind radio surveys have become available, some optical surveys have 
aimed to detect the H$\alpha$ emissions from the HI-selected galaxies. For example, the 
Survey for Ionization in Neutral Gas Galaxies (SINGG) focuses on detecting the H$\alpha$ 
emissions from the HIPASS galaxies (e.g., Meurer et al. 2006, and references therein). 
The SFRs of $\sim 90$ galaxies are successfully estimated by the H$\alpha$ images of the 
HIPASS samples. 

In Figure 7(b), we plot a contour map between the SFR and the HI mass in galaxies for 
the LS model. The SFRs range widely from $10^{-4}$ to $10^{2}$ $M_{\odot}$ yr$^{-1}$ in 
the mass range $10^{7} \la M_{\rm HI}/M_{\odot} \la 10^{10.5}$. Together with the SFRs 
based on the SINGG (filled circles; Meurer et al. 2006), we also show the SFRs 
determined from the intensities of infrared (red dotted box) and radio (green 
dotted box) emissions from the HIPASS galaxies: HIPASS Optical Catalogue (HOPCAT; Doyle 
\& Drinkwater 2006). Furthermore, the SFRs in low surface brightness galaxies based on 
the optical and infrared imaging data are also plotted individually (open circles,  
van Zee et al. 1997; filled triangles, O'Neil, Oey, \& Bothun 2007; and filled 
squares, Rahman et al. 2007).  
Our result is consistent with the observations because the scatter is similar to those 
in the observations for the wide range of the HI masses. 
At low-mass end $M_{\rm HI} \sim 10^{7}$ $M_{\odot}$, 
our model would be constrained more 
stringently by further observational measurements of the SFRs in less massive galaxies. 
This contour map also shows that 
the scatter of the SFR is generally smaller than those in the $J$-band luminosity.
We conclude that, {\it  including the less massive galaxies 
($M_{\rm HI} \sim 10^{7} M_{\odot}$), 
the HI mass correlates with the SFR,  e.g., the $H \alpha$ luminosity 
more tightly than the $J$-band luminosity}. We also show the result of a direct measurement 
of SFR, $\sim 10^{-2.2} M_{\odot}$ yr$^{-1}$, in the nearest galaxy 
($M_{\rm HI}\sim 10^{9.2} M_{\odot}$ at $z=0.009$) giving rise to a damped Ly$\alpha$ 
absorption line (cross; Schulte-Ladbeck et al. 2004). We find that our prediction 
is consistent with the SFR in the DLA galaxy that is also within the range of 
the SFRs obtained from the HIPASS sample.  

The $J$-band luminosities in the $M_{\rm HI}$-$L_{\rm J}$ planes (Figures 5 and 6)
show scatters larger than the SFRs in the SFR-$M_{\rm HI}$ plane (Figure 7(b)).  
This stems from the fact that the $J$-band luminosity is sensitive to the {\it stellar mass} rather 
than the HI mass. Galaxies have a variety of stellar masses even if they have the same amount of HI gas. 
Therefore, the variety of the stellar masses produces  
substantial scatters of the $J$-band luminosity, as shown in the 
$M_{\rm HI}$-$L_{\rm J}$ planes. Furthermore, 
the $B$-band luminosity correlates with the HI mass more tightly than 
the $J$-band (Figures 5(c) and 6(c)) because 
the $B$-band luminosity is more sensitive to the {\it SFR} than the $J$-band 
luminosity.

Figure 8(a) shows the distribution functions of galaxies as a function of the logarithmic SFR
 for the HS model (solid line) and the LS model (dashed line) .  
It is obvious that the SFRs range 
widely from $10^{-6}$ to $10^{2} M_{\odot}$ yr$^{-1}$. We find the mean values 
of the logarithmic SFRs $\langle \log$ SFR [$M_{\odot}$ yr$^{-1}] \rangle$ $\sim -3.2$
for the HS model and $\sim -3.5$ for the LS model. For the LS model, 
the number fractions of galaxies with low SFRs (e.g., $\sim 10^{-4}$ 
$M_{\odot}$ yr$^{-1}$) are higher than those for the HS model, because the star formation 
timescales for the LS model are relatively long in less massive systems. This result suggests 
that less massive galaxies with low SFRs ($\la 10^{-2}$ $M_{\odot}$ yr$^{-1}$) are 
numerically dominant. 

Figure 8(b) shows the distribution functions of galaxies as a function 
of the logarithmic SFR per unit area for the HS model (solid line)  and 
the LS model (dashed line).  The difference between the two models 
is clearer than that shown in Figure 8(a). The LS model obviously predicts 
smaller SFRs per unit area than the HS model. The mean values of 
the logarithmic SFRs per unit area in units of $M_{\odot}$ yr$^{-1}$ kpc$^{-2}$ 
are $-3.43$ for the HS model and $-3.74$ for the LS model. 
When the large sample of SFRs per unit area in the HI-selected galaxies 
becomes available, in addition to the SFRs, the SFRs per unit area 
are a good probe of star formation activities.

\section{HI-selected Galaxies and Sub-DLA systems}
 
We focus on HI-rich systems that contribute to the galaxy population at the low-mass 
end and make further implications concerning the relationship between the HI-selected 
galaxies and quasar absorption systems. In general, quasar absorption systems optically thick 
against ionizing photons are classified into three morphological types: DLA system 
($N_{\rm HI} \ge 10^{20.3}$ cm$^{-2}$), sub-DLA system ($10^{19} < N_{\rm HI}  
< 10^{20.3}$ cm$^{-2}$), and Lyman-limit system ($N_{\rm HI} \ga 10^{17}$ cm$^{-2}$). 
The sub-DLA system gives rise to HI absorption lines with strong damping wings in quasar 
spectra in a manner similar to DLA systems. From aspects of the study of physical 
and chemical processes in gas optically thick against the ionizing radiation background, 
sub-DLA systems are also of particular interest. In recent studies, the basic properties 
of sub-DLA systems have been explored in comparison with those of DLA systems 
(e.g., P{$\rm \acute{e}$}roux et al. 2003, 2005, 2006, 2007; 
Dessauges-Zavadsky et al. 2003; Christensen et al. 2005; Tripp et al. 2005; 
Briggs \& Barnes 2006; Prochaska et al. 2006; York et al. 2006; Khare et al. 2007; 
Kulkarni et al. 2007; O'Meara et al. 2007; Meiring et al. 2007, 2008). 
These studies offer several interesting 
findings; e.g., (1) sub-DLA systems are more abundant than DLA systems.  
The number density of sub-DLA systems observed in each column-density bin is obviously 
higher than that of DLA systems at a given redshift; (2) the metallicities of 
sub-DLA systems are higher than those of DLA systems; and (3) the metallicity evolution 
is positive: an increase in metallicity with decreasing 
redshift, from $1/100 Z_{\odot}$ at $z \sim 4.5$ to $\sim 1/3 Z_{\odot}$ at 
$z \sim 0.5$ ( P{$\rm \acute{e}$}roux et al. 2003, 2007).

Here, we argue how the galaxies, in which the HI column densities are as high as 
those of sub-DLA systems, contribute to the galaxy population at redshift $z=0$. 
In Figure 4, we showed the number fractions of DLA galaxies relative to galaxies. 
Our result shows that the number fractions of DLA galaxies obviously decrease toward 
the low-mass end. This indicates that another population contributes to the local galaxy
population. We refer to galaxies having gaseous disks that exhibit HI column densities, 
$10^{19} < N_{\rm HI}  < 10^{20.3}$ cm$^{-2}$,  corresponding to those of 
sub-DLA systems as {\it sub-DLA galaxies}. 
In our model, the size of a sub-DLA galaxy is given by the radius $R$ where  
$N_{\rm HI} = 10^{19}$ cm$^{-2}$. The HI column density of a sub-DLA galaxy is 
defined as the HI column density averaged over the radius within $R$. 
It has been argued that HI gaseous disks are truncated at a  
column density of $\sim 10^{19}$ cm$^{-2}$ (e.g., Maloney 1993; Corbelli and 
Salpter 1993). The truncated column density depends roughly on the number density 
$n$ and the flux of the ionizing photons $\phi_{\rm i}$. It is simply estimated as 
$N_{\rm th} \sim 8 \times 10^{18}$ cm$^{-2}$ ($\phi_{\rm i}/10^{-4}$ cm$^{-2}$ 
s$^{-1}$)($n/10^{-2}$ cm$^{-3}$)$^{-1}$ under a gas temperature of $10^{4}$ K, 
which is consistent with the column densities of sub-DLA systems. 
Therefore, sub-DLA galaxies having  the gaseous disks with $N_{\rm HI} \sim 10^{19}$ 
cm$^{-2}$ provide us with useful information about the physical condition for 
gas in the outer part of galactic disks and for surroundings of disks.

\subsection{Number Fraction} 

In Figure 9(a), we show the number fractions of sub-DLA galaxies relative to galaxies 
per bin of HI mass for the HS model (solid line) and the LS model 
(dashed line). The number fraction at each bin of HI mass is defined as a ratio of 
the number density of sub-DLA galaxies relative to that of galaxies per bin of HI mass.
In contrast to DLA galaxies, the number fractions of sub-DLA galaxies increase toward 
the low-mass end, particularly for the HS model.  Sub-DLA galaxies are dominant at the low-mass 
end while DLA galaxies are  dominant at  the high-mass end. For the HS model the number 
fractions attain $60 \%$ at $M_{\rm HI} \sim 10^{8} M_{\odot}$ and $80 \%$ at 
$M_{\rm HI} \sim 10^{7} M_{\odot}$, respectively. For sub-DLA galaxies, we find that 
the mean HI gas mass $\langle M_{\rm HI} \rangle$  $= 1.7 \times 10^{8} M_{\odot}$,  
the mean logarithmic HI-mass $\langle \log (M_{\rm HI}/M_{\odot}) \rangle = 6.4$,  
and the cross section weighted mean HI mass $\langle M_{\rm HI} \rangle$  
$= 7.8 \times 10^{9} M_{\odot}$ , while for DLA galaxies   
$\langle M_{\rm HI} \rangle = 5.6 \times 10^{8} M_{\odot}$,   
$\langle \log (M_{\rm HI}/M_{\odot}) \rangle= 7.7$,  
 and the cross section weighted  $\langle M_{\rm HI} \rangle$  
$= 8.2 \times 10^{9} M_{\odot}$ .  
The results indicate that sub-DLA galaxies are primarily composed of less massive galaxies 
rather than DLA galaxies. It should be noted that, in our result, some systems 
are identified not only as DLA galaxies but also as sub-DLA  galaxies 
because the gaseous disks having central column densities higher than the DLA 
threshold of $10^{20.3}$ cm$^{-2}$ extend to the edge at which $N_{\rm HI}$ is 
as low as the sub-DLA threshold of $10^{19}$ cm$^{-2}$.  This causes a case where, in 
a bin of HI mass, the sum of the number fractions for DLA and sub-DLA galaxies is not unity 
(Figures 4 and 9( a)). 
For the LS model, at $M_{\rm HI} \la 10^{9} M_{\odot}$, the number fractions of sub-DLA 
galaxies are lower than those for the HS model, $ \sim 40 \%$ at 
$M_{\rm HI} \sim 10^{8} M_{\odot}$ and $ \sim 30 \%$ at $M_{\rm HI} \sim 10^{7} M_{\odot}$.  
The LS model predicts, for sub-DLA galaxies, the mean HI gas mass $\langle M_{\rm HI} \rangle$  
$= 2.2 \times 10^{8} M_{\odot}$, the mean logarithmic HI-mass 
$\langle \log (M_{\rm HI}/M_{\odot}) \rangle = 7.3$,  
 and the cross section weighted mean HI mass $\langle M_{\rm HI} \rangle$  
$= 6.3 \times 10^{9} M_{\odot}$ ,  
while, for DLA galaxies,  $\langle M_{\rm HI} \rangle = 2.9 \times 10^{8} M_{\odot}$,   
$\langle \log (M_{\rm HI}/M_{\odot}) \rangle= 7.4$,  
and the cross section weighted  $\langle M_{\rm HI} \rangle$  
$= 4.3 \times 10^{9} M_{\odot}$ . 

The main reason for the difference between the number fractions for the models is that, 
for the LS model, the SFRs are relatively low in less massive systems. As shown 
in Figure 8(b), the high SFRs for the HS model result in higher surface densities of stars 
compared to the LS model. This means that the consumption of neutral gas by star formation 
result in a decrease in the HI-column density. This, in turn,  increases the number fraction of 
sub-DLA galaxies particularly at the low-mass end. We want to emphasize that, even for the LS model, 
the number fractions are not negligible for sub-DLA galaxies, i.e., $30\%-40 \%$ for each 
bin of HI mass ($10^{7} \la $ $M_{\rm HI}/M_{\odot}$ $ \la 10^{8}$).     

The number fraction may be greatly dependent on the suppression mechanism of forming 
galaxies embedded in less massive halos. In our model, the isolated halo must have 
$V_{\rm circ}$ larger than the threshold, $V_{\rm circ,th}$, which corresponds to the 
effective Jeans scale. In Figure 9(b), for the HS model, we present  number fractions of 
DLA galaxies (dash-dotted line) and sub-DLA galaxies  (dotted line) when 
$V_{\rm circ,th}=40$ km s$^{-1}$, together with   those of DLA galaxies ({\it solid line}) 
and sub-DLA galaxies  (dashed line) when $V_{\rm circ,th}=30$ km s$^{-1}$. 
Because $V_{\rm circ,th}$ provides the lowest mass of dark halos, adopting a small 
$V_{\rm circ,th}$ generally prompts the formation of less massive systems.
However, we find that the number fractions show little difference for the 
cases adopting $V_{\rm circ,th}=40$ km s$^{-1}$ and $30$ km s$^{-1}$. 
This means that the number fractions are almost independent of the suppression threshold. 
Adopting large $V_{\rm circ,th}$ results in reducing the number of galaxies and 
DLA galaxies simultaneously, and thus $V_{\rm circ,th}$ does not affect the number fractions.  
This is also the case for the sub-DLA galaxies. 
These results suggest that the uncertainties in 
the threshold on halo circular velocity or the ionization history in the universe 
do not change our results.

The majority of HI-selected galaxies detected in the {\it latest} radio surveys have 
gaseous disks where the HI column densities are above the DLA threshold. 
However, our results suggest that gaseous disks of 
less massive galaxies ($M_{\rm HI} \la 10^{8} M_{\odot}$) have  
 HI column densities above  the {\it sub-DLA } threshold. 
When the observations have sufficiently lower detection limits 
 ($M_{\rm HI} \la 10^{8} M_{\odot}$), 
the observed populations will switch from being dominated by DLA galaxies to being 
dominated by sub-DLA galaxies.  Sub-DLA galaxies will provide us with stringent 
constraints on formation and evolution processes of galaxies.

\subsection{Cross Section}

Here, we argue how sub-DLA galaxies are detected in radio observations. 
In Figure 10(a), we present a contour map for the cross sections of DLA 
galaxies as a function of the HI mass for the HS model. 
Obviously, the cross sections correlate strongly to the HI masses, and  
the disk sizes of less massive galaxies are small; 
$\sim 3 h^{-1}$ kpc at $M_{\rm HI} \sim 10^{8} M_{\odot}$ and $\sim 1 h^{-1}$ kpc 
at $M_{\rm HI} \sim 10^{7} M_{\odot}$. 
In the observation, \citet{RS03} investigated cross sections of DLA galaxies on the 
basis of the HI-selected samples including the ADBS ($\sim 100$ galaxies). 
The observational data are also plotted as {\it dots} in Figure 10(a). 
Our result is entirely consistent with the observation because the 
$\sigma$-$M_{\rm HI}$ contours 
obviously trace  the observations in the range of HI mass 
$10^{7} \la M_{\rm HI}/M_{\odot} \la 10^{10.5}$.

In Figure 10(b), we plot the mean  logarithmic cross sections of DLA galaxies 
with one $\sigma$ error bars for the HS model (red line). 
It is evident that the logarithmic cross section, $\log \sigma$, is linearly proportional to 
the logarithmic HI mass, $\log M_{\rm HI}$; $\sigma$ $\propto$ $ M_{\rm HI}^{\alpha}$. 
For the HS model, we find $\alpha=0.97 \pm 0.01$ (red line) at 
$10^{6} \le M_{\rm HI} / M_{\odot} \le 10^{10.5}$ when the relation is fitted by 
averaged least squares.  
In Figure 10(b),  for comparison to the result for the HS model, the mean cross sections 
of DLA galaxies for the LS model  is also plotted (blue line). For the LS model, the 
$\sigma$-$M_{\rm HI}$ relation has a slope of $\alpha=0.98 \pm 0.01$ (blue line). 
We find that there is no significant difference between the 
$\sigma$-$M_{\rm HI}$ relations predicted by the two models.

In Figure 10(b), we present a contour map for the cross sections of sub-DLA galaxies  
as a function of HI mass for the HS model. Our result shows that the cross section, 
$\sigma$, tightly correlates to the HI mass. In the range of the HI mass 
$10^{6} \le M_{\rm HI} / M_{\odot} \le 10^{10.5}$, we find the slopes 
$\alpha= 0.92 \pm 0.01$ ( $\sigma$ $\propto$ $M_{\rm HI}^{\alpha}$) for the HS model  
and $0.96 \pm 0.01$ for the LS model when the relation is fitted by averaged least squares. 
We find that the slope of the $\sigma$-$M_{\rm HI}$ relation is insensitive to the SFRs in the disks.
For comparison to the $\sigma$-$M_{\rm HI}$ relation for DLA galaxies (red line), 
sub-DLA galaxies have cross sections obviously larger than DLA galaxies because the disk radius 
at  the sub-DLA limit $N_{\rm HI} = 10^{19}$ cm$^{-2}$ is typically larger than that at the 
DLA limit $N_{\rm HI} = 10^{20.3}$ cm$^{-2}$ under our assumption that the radial 
distribution of HI column density follows an exponential profile. The HI column densities are relatively 
low in the outer part of a gaseous disk whereas the HI column densities are high in the inner part. 
This means that the outer parts (large cross sections) of disks correspond to sub-DLAs 
while the inner parts (small cross-sections) to DLAs.

\subsection{Size}

In Figure 11, we present the distribution functions of (a) DLA galaxies, 
(b) sub-DLA galaxies, and (c) galaxies as a function of the disk radii for the HS 
model (solid line) and the LS model (dashed~line). 
The number fraction is defined as a ratio of the number within each bin of a radius $b$ 
relative to the total number. Figure 11(a) shows that the number fraction increases 
toward small radii. The mean disk radii of DLA galaxies are $2.7 h^{-1}$ kpc for the HS 
model and $1.7 h^{-1}$ kpc for the LS model. The LS model predicts 
a mean radius of gaseous disks smaller than the HS model because, for the LS model, 
less massive galaxies with small radii contribute to the population of DLA galaxies  
(Figure 4).  By contrast, at large radii, the number fractions for the HS model are 
higher than those for the LS model.

In Figure 11(b), the distribution functions of {\it sub-DLA galaxies} 
are similar to those of DLA galaxies. 
We find the mean radii of gaseous disks,  $3.5 $ $h^{-1}$ kpc for the HS 
model and $4.4 $ $h^{-1}$ kpc for the LS model. 
Conversely to the results for DLA galaxies, the HS model produces a mean disk radius 
smaller than that for the LS model. This is because, for the HS model, less massive galaxies 
($M_{\rm HI} \la 10^{8} M_{\odot}$) with small disk radii  contribute 
primarily to the sub-DLA population (Figure 9). 
 
For both models,  the mean radii of sub-DLA galaxies are larger than those of DLA galaxies
because a truncated column density for sub-DLA galaxies smaller than that for 
DLA galaxies results in the mean disk radii of sub-DLA galaxies larger than those of DLA galaxies. 
This result suggests that some massive $L^{*}$ galaxies with large 
disks could also be identified as host galaxies of sub-DLA systems rather than DLA systems 
when a line of sight to the background bright source is located at the disk edge where 
the HI column density is $\sim 10^{19}$ cm$^{-2}$.

Zwaan et al. (2005b) investigated the properties of DLA galaxies at $z=0$ 
in comparison to those of DLA systems at $z>0$ on the basis of an HIPASS sample of 
$355$ radio emission line maps using the Westerbrock Synthesis Radio Telescope. 
They conclude that DLA systems arise in galactic disks because the basic observational
quantities of DLA galaxies are consistent with those of DLA systems obtained from the 
optical and UV surveys. They also argue which morphological types of galaxies have HI 
column densities as high as DLA systems on the basis of their statistical analysis of 
the galaxy population at redshift $z=0$. 
For example, they find that the median impact parameter between 
the line of sight to a background quasar and the galactic center is $5.5 h^{-1}$ kpc, 
which is larger than the typical value of $3.0 h^{-1}$ kpc for the HS model in our calculation. 
This discrepancy stems from our model taking into account less massive systems below the 
observational detection limit. Our result suggests that less massive systems 
($M_{\rm HI} \la 10^{8} M_{\odot}$) partly contribute to the population of DLA galaxy  
at $z=0$ (Figure 4). For DLA galaxies with the HI mass $M_{\rm HI} \ge 10^{8} M_{\odot}$, 
we find the average values $5.2 h^{-1}$ kpc (the HS model), consistent with the observation. 
Therefore, we want to emphasize that, although the number fraction of DLA galaxies to galaxies 
is small in the low-mass range  ($M_{\rm HI} \la 10^{8} M_{\odot}$), the contribution 
of less massive galaxies to DLA galaxies is not negligible. 

 Figure 11(c) is similar to Figure 11(a) but for galaxies.  
The radius is defined as that at which 
$N_{\rm HI}=10^{19}$ cm$^{-2}$ given by ionization equilibrium against the UV 
background radiation mentioned above. It should be noted again that 
there is no criterion when picking out `galaxies' while 
the sub-DLA galaxies are picked out 
by a condition on the central HI column density of 
$10^{19} \leq N_{\rm HI} \leq 10^{20.3}$ cm$^{-2}$.  
The number fraction of galactic radii increases toward small radii, in a 
manner similar to those of DLA and sub-DLA galaxies. We find that there is little 
difference in the distribution between the HS and the LS models. The mean radii are 
$1.6 $ $h^{-1}$ kpc for the HS model and $1.4 $ $h^{-1}$ kpc for the LS model. 
This indicates that the difference between the SFRs does 
not significantly affect the number fractions. The result shows that the accumulated 
number fractions at $b \geq 3$ $h^{-1}$ kpc attain $\sim 17 \%$ for the HS and the LS 
models, and those at $b \geq 1$ $h^{-1}$ kpc attain $\sim 60 \%$ for the HS model and 
$\sim 56 \%$ for the LS model.    

\subsection{Star Formation Rate}

Figure 12(a) shows the distribution functions of DLA galaxies as a function of 
the logarithmic SFR 
for the HS model (solid line) and the LS model (dashed line). 
The SFRs in DLA galaxies widely range from $10^{-6}$ to $10^{2}$ 
$M_{\odot}$ yr$^{-1}$ similar to those in galaxies shown in Figure 8(a). 
 We find the mean values of the logarithmic SFRs: $ \langle \log$(SFR)
[$M_{\odot}$ yr$^{-1}] \rangle$ $\sim -2.3$ for the HS model and $\sim -3.7$ for 
the LS model. In comparison to the result  (Figure 8(a)) for galaxies without any 
threshold on HI column density of gaseous disk, it appears that 
in the range of the SFRs $\la 10^{-4}$ $M_{\odot}$ yr$^{-1}$ the HS model predicts 
the smaller number fractions than those of galaxies. 
This suggests that the galaxies with low SFRs do not significantly contribute to DLA 
galaxies but they do for sub-DLA galaxies.

Figure 12(b) shows the distribution functions of sub-DLA galaxies as a function of 
the logarithmic SFR 
for the HS model (solid line) and the LS model (dashed line). 
We find that the mean values of the logarithmic SFRs, $\langle \log$(SFR) 
[$M_{\odot}$ yr$^{-1}] \rangle$ $\sim -3.7$ for the HS and the LS models. 
The mean values show little difference between the two models. 
 For the HS model, sub-DLA galaxies are composed primarily of 
galaxies with low SFRs ($\sim 10^{-4} M_{\odot}$ yr$^{-1}$), 
which is in contrast with the result 
for DLA galaxies where the SFRs are high ($\sim 10^{-2} M_{\odot}$ yr$^{-1}$). 
This stems from the fact that sub-DLA galaxies consist mainly of the less massive galaxies 
while DLA systems consist mainly of massive ones (Figure 9(a)). We also find that 
there is no significant difference between the number fractions of sub-DLA galaxies and 
galaxies in the range of the SFRs, $\la 10^{-2} M_{\odot}$ yr$^{-1}$ (Figure 8(a)).  

\subsection{Luminosity}

In Figure 13, we show the $B$-band luminosity distribution as a function of HI mass of 
sub-DLA galaxies for (a) the HS model and (b) the LS model. The average 
luminosities with  $1 \sigma$ error bars are shown as {\it solid lines}. Our results show 
that massive sub-DLA galaxies are luminous in the $B$-band. We find that sub-DLA galaxies 
have the optical luminosities that range widely up to $\sim 10^{11} L_{\odot}$, with the 
mean values of the $B$ band luminosities $\langle L_{\rm B} \rangle$ 
$\sim 4.2 \times 10^{8} L_{\odot}$ for the HS model and $\sim 3.6 \times 10^{8} L_{\odot}$ 
for the LS model. The mean luminosities in the $B$ band differ little because the mean SFR 
for the LS model matches that for the HS model (Figure 12(b)). In the 
$L_{\rm B}$-$M_{\rm HI}$  plane, the luminosity distributions of sub-DLA galaxies are clearly 
different for the two models. For the LS model, sub-DLA galaxies with low HI-mass exhibit low  
$B$-band luminosities down to $L_{\rm B} \sim 10^{6} L_{\odot}$ while for the HS model 
there are no sub-DLA galaxies with luminosities $L_{\rm B} \sim 10^{6} L_{\odot}$. 
This difference is attributable to the fact that, for the LS model, the number fractions of sub-DLA 
galaxies with low SFRs are larger than those for the HS model (Figure 12(b)). 
In addition to the H$\alpha$ luminosities, the $B$-band luminosities of sub-DLA galaxies 
are a good probe to investigate the SFRs in less massive galaxies. Together with  
previous results, we conclude that sub-DLA galaxies, which typically comprise less massive 
systems ($\langle M_{\rm HI} \rangle = 2 \times 10^{8} M_{\odot}$ and 
$\langle \log (M_{\rm HI}/M_{\odot}) \rangle \sim 7.5$),   
are compact ($\langle b \rangle \sim 4 h^{-1}$ kpc) and 
optically faint ($\langle L_{\rm B} \rangle  \sim 4 \times 10^{8} L_{\odot}$).

\section{Discussion}

\subsection{HI Mass Functions} 

Our models produce a good match to the HIMFs given by various blind radio surveys 
as shown in Figure 2. However, it appears that  the HS model slightly underpredicts the 
number densities at the low-mass end. A different model that adopts a different threshold 
for the halo circular velocity of less than $V_{\rm circ,th} = 30$ km s$^{-1}$ may 
provide better agreement  with the observations.  
 
As argued in Section 5, the suppression of forming dwarf galaxies in halos with small 
circular velocities may be caused by the ionizing background radiation. Because the 
radiation intensity remains uncertain at low redshift, this may allow to adopt a threshold 
smaller than $V_{\rm circ,th} = 30$ km s$^{-1}$ for calculating the HIMF. 
The circular velocity threshold $V_{\rm circ, th}$, which gives the lowest circular 
velocity for galactic halos, sensitively depends on the intensity of UV radiation and the 
reionization process. Thus, the threshold $V_{\rm circ,th}$ should affect the  number of 
less massive galaxies and the shape of HIMFs. For example,  Somerville (2002) 
investigated the effect of suppressing the formation of less massive galaxies due to 
the squelching of gas infall by the ionizing background radiation in order to resolve the 
excess number of dwarf galaxies predicted by a semi-analytic model. 
They find that halos with $V_{\rm circ,th} < 30$ km s$^{-1}$  are not able to 
accrete  gas. This suggests that photoionization squelching suppresses formation 
of less massive galaxies in halos with $V_{\rm circ,th} < 30$ km s$^{-1}$. 
However, Okamoto et al. (2008) argued that,  in Somerville (2002), the fitting  
function, concerning accreted gas in a halo,  overestimates the mass of the halo within 
which gas can cool.  On the basis of the result obtained by the numerical simulation, 
the authors require the threshold  $V_{\rm circ,th}$  to be less than $20$ km s$^{-1}$.
For the HS model in our calculation, adopting a small threshold of  $V_{\rm circ,th}$   
$< 30$ km s$^{-1}$ may produce an increase of less massive galaxies and give better 
agreement with the observed abundance at the low-mass end of HIMFs. 

The underestimate of HIMFs at low-mass end might also be caused by the shape of 
the adopted halo mass function used here. In this calculation, we adopt a mass function 
of dark halo progenitors given by an extended Press--Schechter formalism. The results 
of the numerical simulations show some discrepancies between the mass function of 
dark halos and those given by the extended Press-Schechter formalism, in particular 
for those  at the low mass end (e.g., Somerville et al. 2000; Yahagi, Nagashima \& Yoshii 
2004; Li et al. 2007). Although no definite conclusion has yet been reached regarding 
which mass functions should be adopted here, this may cause an underestimation of  
the number densities of less massive galaxies at $M_{\rm HI} \la 10^{7} M_{\odot}$. 

It should be noted that there are uncertainties in the determination of the HIMF slope from 
observations. The low-mass end of the HIMF suffers from statistical uncertainties due to the 
small sample size. For example, the ADBS sample includes only a dozen galaxies and the 
HIPASS does $\sim 40$ galaxies with $M_{\rm HI} < 10^{8} M_{\odot}$. 
Additionally, there are also uncertainties in the distance estimation \citep{MHG04}. 
However, the HIMF slope should place stringent constraints on the formation of dwarf galaxies 
as well as the optical luminosity function. This is a clear point that deserves further investigation. 

It is also valuable to note the total HI mass density parameter, that is the HI mass density of 
galaxies as a fraction of the cosmic critical density. We find $\Omega_{\rm HI}=5.55$ 
$\times$ $10^{-4}$ for the HS model and $1.04$ $\times$ $10^{-3}$  for the LS model. 
In comparison to the observations,  we find that the HI mass density parameter for the HS model 
is consistent with the HIPASS measurement $\Omega_{\rm HI}=3.5 \pm 0.4 \pm 0.4$ 
$\times$ $10^{-4}$ \citep{Z05a}.

\subsection{Number Densities}

Here we address the number density per unit redshift $dN/dz$. Our results present 
$dN/dz$ of DLA galaxies $\sim 4.0 \times 10^{-2}$ at $z=0$ for the HS model and 
$6.3 \times 10^{-2}$ for the LS model. The LS model produces larger $dN/dz$ than 
the HS model because, for the LS model the low SFRs in DLA galaxies cause high HI 
column densities above the DLA threshold. In observations, the ADBS sample for DLA 
galaxies at $z=0$ yields $dN/dz = (5.3 \pm 1.3) \times 10^{-2}$  \citep{RS03},  
the HIPASS $ \sim (5.8\pm 0.6 ) \times 10^{-2}$ \citep{Ryan03} and 
 $(4.5 \pm 0.6) \times 10^{-2}$ \citep{Z05b}. These are consistent with our 
results.  
Because our models reproduce the $dN/dz$ and cross sections, our result suggests 
that, even in the case that less-massive DLA galaxies 
($M_{\rm HI}\sim 10^{7} M_{\odot}$) are taken into account,  DLA systems 
at $z=0$ arise primarily in galactic disks rather than tidal gas caused by galaxy interactions 
or gas driven by galactic winds.  

Furthermore, our models predict $dN/dz$ of sub-DLA galaxies: $dN/dz \sim 0.15$ for 
the HS model and $\sim 0.17$ for the LS model. These are roughly $2-3$ times larger 
than those of DLA galaxies. We also find that there is no difference between the sub-DLA 
$dN/dz$ for the two models. The reason is considered as follows. At a given redshift, the 
$dN/dz$ is described as $dN/dz \propto f n_{\rm gal} \sigma$, where $f$ is the number 
fraction of sub-DLA galaxies to galaxies, $n_{\rm gal}$ is the number density of galaxies, 
and $\sigma$ is the cross section of the galactic disk. We showed that there is little difference 
in the cross-sections $\sigma$ between the models. 
However, for the HS model, the number fractions of sub-DLAs $f$ are larger than those for 
the LS model, while the LS model produces the number densities $n_{gal}$ higher than those of 
the HS model. This results in no significant difference between the values $f n_{\rm gal}$ 
and $dN/dz$ for the two models. Our result is also consistent with an estimation of 
$dN/dz = 0.21$ at  redshift $z < 2$ on the basis of an analysis of the HI column density 
distributions of high-redshift sample \citep{Peroux05}.  

\subsection{Further Implications}

Many ideas concerning the nature of DLA systems have been proposed through  
observational/theoretical studies: large, massive spiral galaxies (e.g., Wolfe et al. 1986), 
gas-rich dwarf galaxies (e.g., York et al. 1986), proto-galactic infalling gas that 
is not virialized (e.g., Haehnelt et al. 1998), and so on.  Although it is not clear which 
scenarios are valid, a dramatic change in the nature of DLAs may occur at redshift $z \sim 2$ 
from the results of the numerical simulations (e.g., Gardner et al. 2001; Nagamine et al. 
2004, 2007) and semi-analytic models (e.g., Kauffmann 1996; Maller et al. 2001; 
Okoshi et al. 2004; Okoshi and Nagashima 2005; Johansson and Efstatiou 2006). 
At redshift $z \sim 0$, the radio observations suggest that compact HI objects 
($M_{\rm HI} \ga 10^{8} M_{\odot}$) mainly exhibit the HI column densities 
comparable to those of DLA systems. They are consistent with our results showing that 
compactly bound systems including less massive ones 
($M_{\rm HI} \sim 10^{7} M_{\odot}$) show the basic properties of DLA systems. 

If the compactly bound objects also comprise sub-DLA galaxies, how can they be  
detected in the observations?  
In future, when the large samples including less massive galaxies 
at $M_{\rm HI} \la 10^{7} M_{\odot}$ become available, it is probable 
to identify them as sub-DLA galaxies rather than DLA ones.  The HI-selected samples have 
an advantage in terms of  detectability of the optical counterparts over those of quasar 
absorption systems, sub-DLA systems. Because the HI-selected samples provided by the 
blind surveys are not behind bright background quasars, they do not suffer from the masking 
effect. No glare of background quasars does not prevent from the detection of weak optical 
emission lines from the HI-selected galaxies if their surface brightness is very low. If the host 
galaxies of sub-DLA systems are compact, the masking effect would be serious in the searches 
particularly for the optical counterparts of sub-DLA systems just as much as DLA systems. 
In the optical surveys for  less massive galaxies ($M_{\rm HI} \sim 10^{7} M_{\odot}$),  
it would be quite valuable to pay attention to some of the difficulties in detecting the sub-DLA 
hosts due to the compactness and the faintness.  

Some observations suggest that a sub-DLA host comprises compact and less massive 
systems ($M_{\rm HI} \sim 10^{7} M_{\odot}$). For example, Tripp et al. (2005) 
explored a nearby sub-DLA system with $\log N_{\rm HI}=19.32 \pm 0.03$ at $z_{abs}=0.006$ 
on the periphery of the Virgo Cluster by using the Space Telescope Imaging Spectrograph on 
board the {\it Hubble Space Telescope}. They argue that they could not find any bright galaxies in the
proximity of the line of sight toward the background quasar PG1216+019. The nearest galaxy is a 
sub-$L^{*}$ galaxy with a projected distance of $\sim 86 h_{\rm 75}^{-1}$ kpc. 
However, a radio observation successfully detected the radio emissions using the Westerbrock 
Synthesis Radio Telescope \citep{BB06}. The emission feature is very compact; it is located 
within $30$ arcsec ($\la 4$ kpc) of the quasar sight line. They argue that, because the
velocity spread of the emission line is between $20$ and $60$ km s$^{-1}$, the radio emission 
feature indicates the HI masses, $5 \times 10^{6} \le M_{\rm HI} / M_{\odot}$ 
$\le 1.5 \times 10^{7}$. The result also suggests that compactly bound objects will be identified 
as the host galaxies of sub-DLA systems. Indeed, it may not be surprising that sub-DLA galaxies 
are very optically faint, consistent with our result, e.g., the mean luminosity in the $B$ band is 
$L_{\rm B} \sim 10^{6.1 \pm 0.8} L_{\odot}$ at $M_{\rm HI} \sim $ 
$10^{7} M_{\odot}$ for the LS model. This observation agrees with our prediction that  sub-DLA 
galaxies with low HI-masses are very faint and compactly bound objects. 

Sub-DLA systems may be associated with high velocity clouds (HVCs). 
Some radio observations show that the HVCs near M31 and M33 have typical HI column densities 
of $10^{19}$-$10^{20}$ cm$^{-2}$, HI masses of $10^{5}$-$10^{6} M_{\odot}$ 
and sizes of $\sim 1$ kpc \citep{WBT05, G08}. On the basis of estimates of the dynamical and 
virial masses, the HVCs are likely gravitationally bounded systems. For HI clouds in the Milky-Way 
halo, some of the absorption features in the spectra toward the background quasars are similar to 
those of intervening quasar absorption systems \citep{Ben08}. 
These properties are consistent with those of sub-DLA systems predicted by our models. 
It is probable that these systems are the sub-DLA hosts. To reveal the sub-DLA hosts,  it needs 
to statistically compare the radio characteristics of the HVCs with the properties of intervening HI 
absorption systems toward quasar/gamma-ray burst. 

Further investigation of the sub-DLA hosts requires a large sample of less massive objects 
($M_{\rm HI} \sim 10^{7} M_{\odot}$), which would be the candidates for 
sub-DLA galaxies. Indeed, a program aimed at obtaining a census of such less massive objects in a 
radio survey is now on going. A recent HI blind survey, the Arecibo Legacy Fast ALFA (ALFALFA) 
Survey, aims to detect on the order of $20,000$ extragalactic HI emission sources out to $z \sim 0.06$ 
including ones with HI masses $M_{\rm HI} < 10^{7.5} M_{\odot}$ by covering $7000$ 
deg$^{2}$ of sky \citep{G05}. This survey is specifically designed to determine the basic properties 
of HI-selected galaxies statistically, including the determination of the HIMF, particularly at the faint end 
($M_{\rm HI} < 10^{8} M_{\odot}$). In conjunction with optical studies, the ALFALFA 
survey will provide us with the basic properties of the optically faint and less massive galaxies including 
several results presented here: the SFR-$M_{\rm HI}$ relation, the relationship between HI cross section 
and $M_{\rm HI}$, and the HI diameter function. The ALFALFA survey is able to map the peripheries of 
HI-selected galaxies to a HI column-density limit on the order of $5 \times 10^{18}$ cm$^{-2}$ that 
corresponds to the threshold of the sub-DLA systems. This means that the survey also provides   
valuable opportunities for exploring the nature of quasar absorption systems, e.g., sub-DLA systems. 
In the near future, comprehensive studies based on large samples of low HI-mass objects will reconcile 
the local galaxy population with the quasar absorption systems.

\section{Conclusion}

We have investigated the properties of HI-selected galaxies in comparison with quasar absorption systems 
at redshift $z=0$ within a hierarchical galaxy formation scenario using a semi-analytic model.  
By drawing a detailed comparison between the properties of the HI-selected galaxies and the HI absorption 
systems, we find that DLA galaxies consist primarily of optically-faint, compactly bound-systems.  
Furthermore, for the purpose of reconciling the quasar absorption system with the galaxy population, we 
have investigated the properties of local HI-selected galaxies. The main conclusions are summarized as follows.\\

1. The SFRs in HI-selected galaxies correlate tightly with the HI masses, SFR 
$\propto$ $M_{\rm HI}^{\alpha}$, $\alpha=1.25-1.40$ in the range of HI mass 
$10^{6} \le M_{\rm HI}/M_{\odot} \le 10^{11}$ (Figure 7).  The SFR ranges widely from 
$10^{-6}$ to $10^{2}$ $M_{\odot}$ yr$^{-1}$ with the mean logarithmic SFR  
$ \langle \log $SFR [$M_{\odot}$ yr$^{-1}] \rangle$ $\sim -3$ (Figure 8). In contrast, the relationship 
between the $J$-band luminosity and the HI mass is quite broad (Figures 5 and 6). \\

2. There is no statistically significant difference between HI mass functions of DLA galaxies and galaxies at 
$M_{\rm HI} \ga 10^{8}$ $M_{\odot}$ (Figure 2). This suggests that DLA galaxies correspond 
primarily to HI-selected galaxies detected in blind radio surveys at 
$M_{\rm HI} \ga 10^{8}$ $M_{\odot}$ (Figure 4).\\

3. Sub-DLA galaxies will replace DLA ones as the dominant population (Figure 9) at the low-mass end 
($M_{\rm HI} \la 10^{8} M_{\odot}$).  The number fractions of sub-DLA galaxies to galaxies are 
between $40\%$ and $60 \%$ at $M_{\rm HI} \sim 10^{8} M_{\odot}$ and between $30\%$ and $80 \%$ at 
$M_{\rm HI} \sim 10^{7} M_{\odot}$. If the detection limits on the HI mass in the blind radio surveys 
are low enough to detect less massive systems ($M_{\rm HI} \sim 10^{7} M_{\odot}$), the
population detected by the surveys will switch to being dominated by sub-DLA galaxies instead of DLA ones. 
In addition to  being a good probe of DLA systems, the HI-selected galaxies can be a strong probe of sub-DLA 
systems. \\

\vspace*{0.5cm}

\acknowledgments

We thank Jessica Rosenberg for kindly providing us with the observational data 
and the referee for a careful reading of this manuscript and for suggestions that 
have improved the clarity of this presentation.  We also thank Alan Hatakeyama  
for improving the paper in English. This calculation is in part carried out 
on the general computer system at the Astronomical Data Analysis Center (ADAC)  
of the National Astronomical Observatory of Japan. This work has been supported 
in part by a Grant-in-Aid for Scientific Research from the Ministry of Education, 
Culture, Sports, Science and Technology (No. 21540245 ).

\newpage

\newpage

\begin{figure}
\plotone{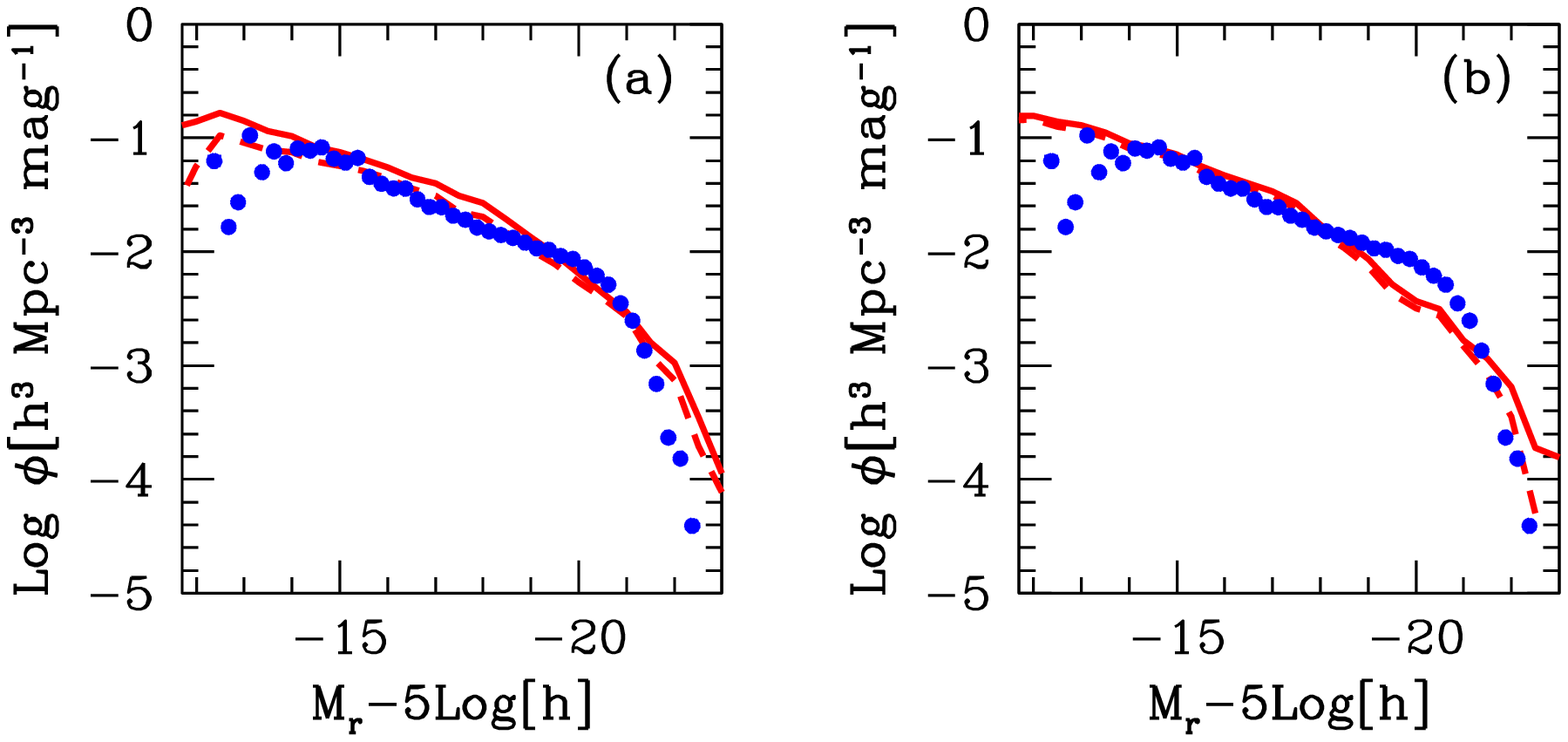}

\caption{
The $r$-band luminosity functions of galaxies ({\it solid line}) and 
DLA galaxies ({\it dashed line}).   
Panel (a) shows the results for the HS model and panel 
(b) shows those for the LS model. 
The observational data from the SDSS DR2  
({\it filled circles}; Blanton et al. 2005) are also shown. 
} \label{fig:corrfil}

\end{figure}

\begin{figure}
\plotone{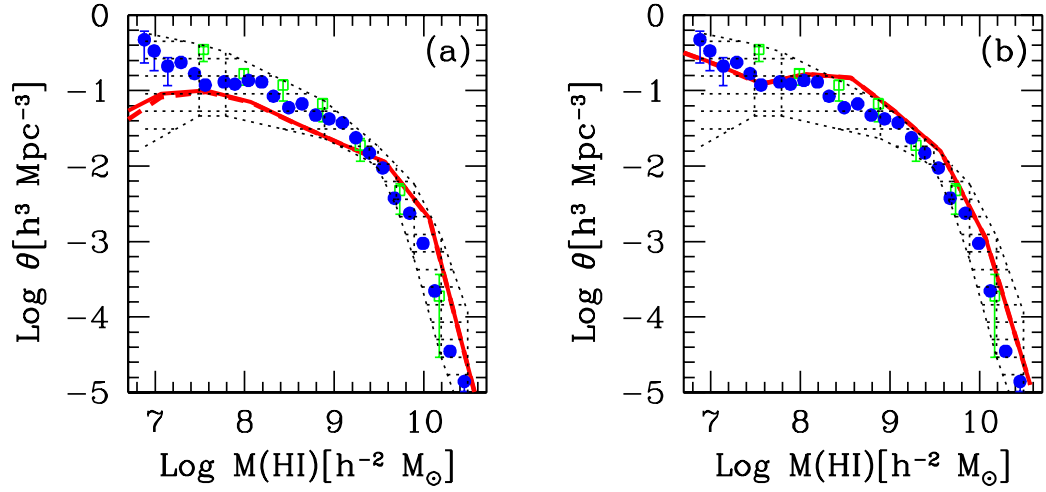}

\caption{
HI mass functions of galaxies ({\it solid line}) 
 and DLA galaxies ({\it dashed line}). 
 Panel (a) shows the results for the HS model and panel (b) shows those for the 
LS model.  The shaded region represents the region of HIMFs given by various blind surveys 
(Henning et al. 2000; Rosenberg \& Schneider 2002, 2003; Springob, Haynes, \& Giovanelli 2005; 
Zwaan et al. 1997, 2003, 2005a). The points are the observational data provided by the ADBS 
({\it open squares}; Rosenberg \& Schneider 2002) and the HIPASS ({\it filled circles}; 
Zwaan et al. 2005a). 
} \label{fig:corrfil}

\end{figure}

\begin{figure}
\plotone{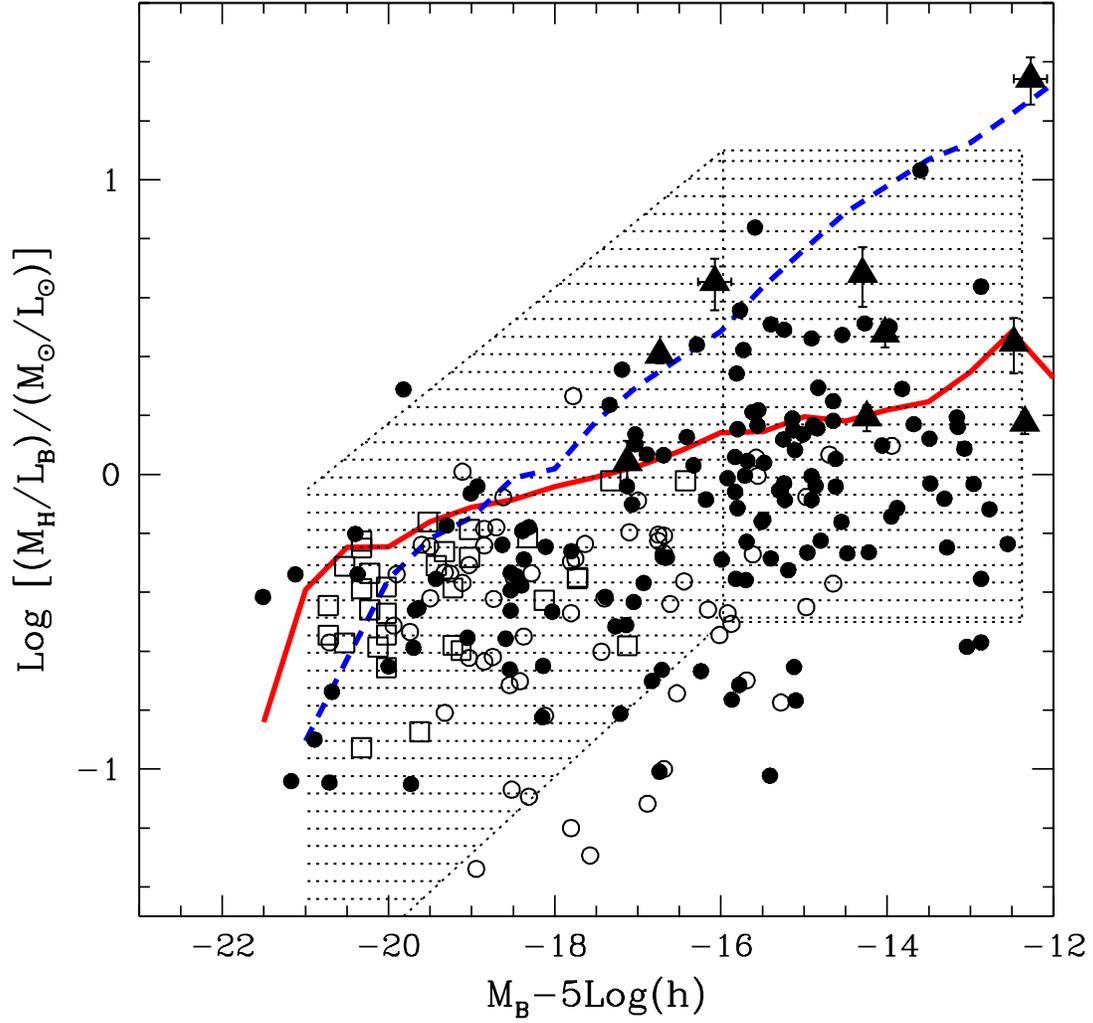}

\caption{Cold-gas masses relative to the $B$-band luminosities of galaxies. The lines 
represent the medians for the HS ({\it solid line}) and LS ({\it dashed line}) models. 
The symbols indicate the observational data given by  Huchtmeier \& Richter (1988; {\it filled circles}), 
Sage (1993; {\it open circles}) and Garnett (2002; {\it open squares}). 
The HI-selected galaxies from the HIPASS  are also included; Koribalski et al. 
(2004; {\it shaded region}) and Warren, Jerjen \& Koribalski (2006; {\it filled triangles}). 
Note that the cold gas mass from the HIPASS and Huchtmeier \& Richter (1988)  
includes only HI.   
} \label{fig:corrfil}

\end{figure}

\begin{figure}
\plotone{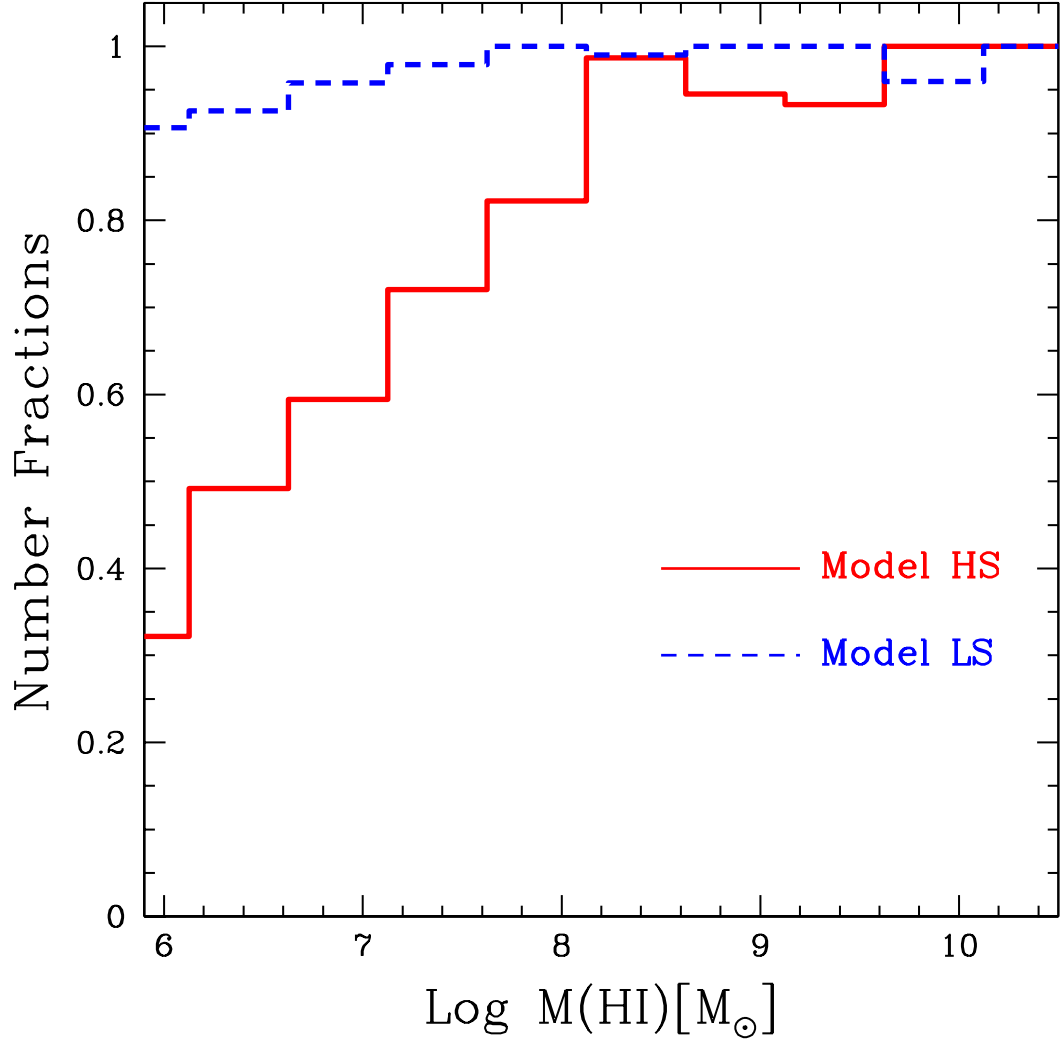}

\caption{Histogram of number fractions of DLA galaxies. The number fraction at each bin 
of HI mass is defined as a ratio of the number density of DLA galaxies to that of galaxies per bin of HI mass.  
The results for the HS and LS models are represented as by solid and  as  dashed lines, 
respectively. ({\it A color version of this figure is available in the online journal. } )
} \label{fig:corrfil}

\end{figure}

\begin{figure*}
\plotone{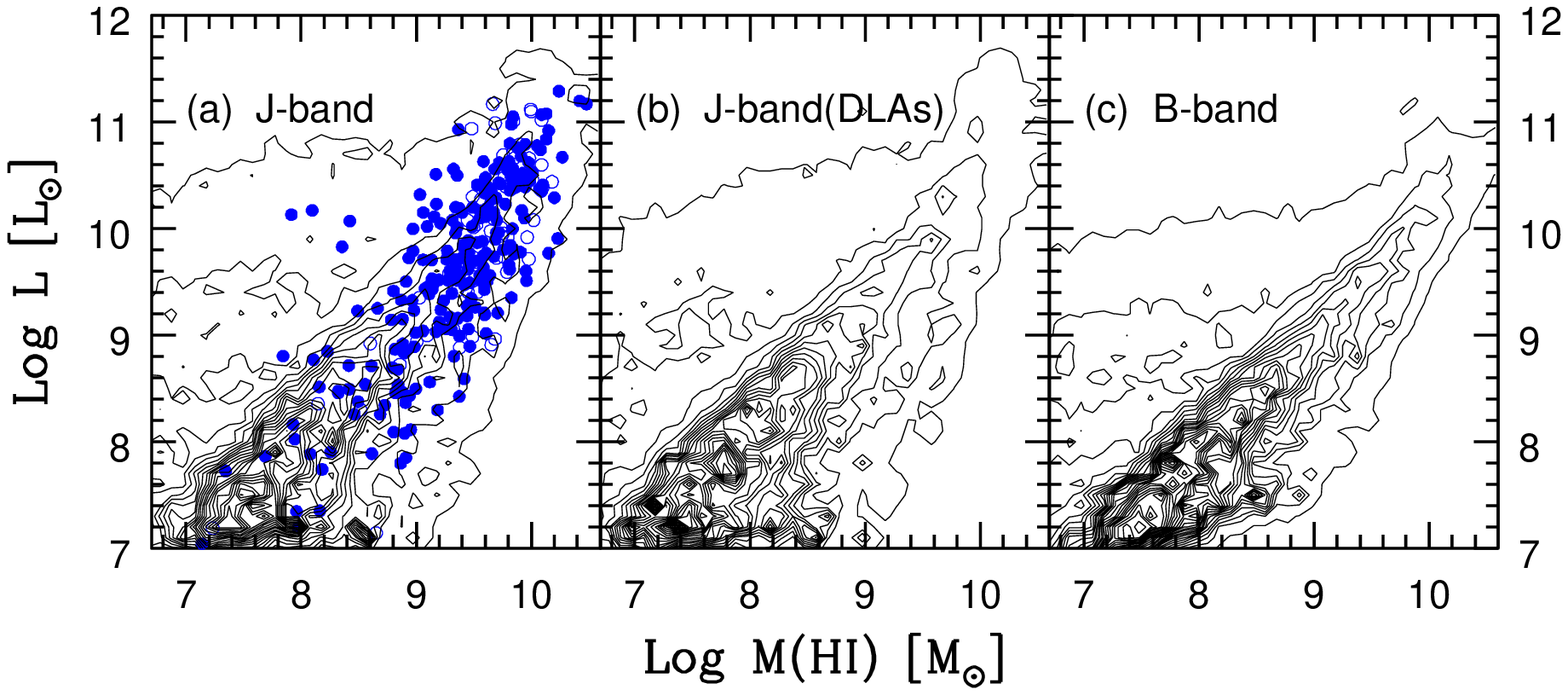}

\caption{
Luminosities of galaxies and DLA galaxies as a function of HI mass at redshift $z=0$ for the 
HS model.  The contour levels are shown in order from outer to inner at an interval of $10^{-4}$ 
based on the calculated number of galaxies and DLA galaxies per logarithm of luminosity per logarithm 
of HI mass. 
(a) The contour map for the HI mass, $M_{\rm HI}$, vs the luminosity in the $J$ band,  
$L_{\rm J}$, of galaxies. The symbols indicate the observational data from the ADBS data 
(filled circles, Rosenberg \& Schneider 2000;  open circles, Spitzak \& Schneider 1998).
(b) The contour map for $M_{\rm HI}$ vs $L_{\rm J}$ of DLA galaxies. 
(c) The contour map for $M_{\rm HI}$ vs the luminosity in the $B$ band $L_{\rm B}$ of  galaxies. 
} 
\label{fig:corrfil}
\end{figure*}

\begin{figure*}
\plotone{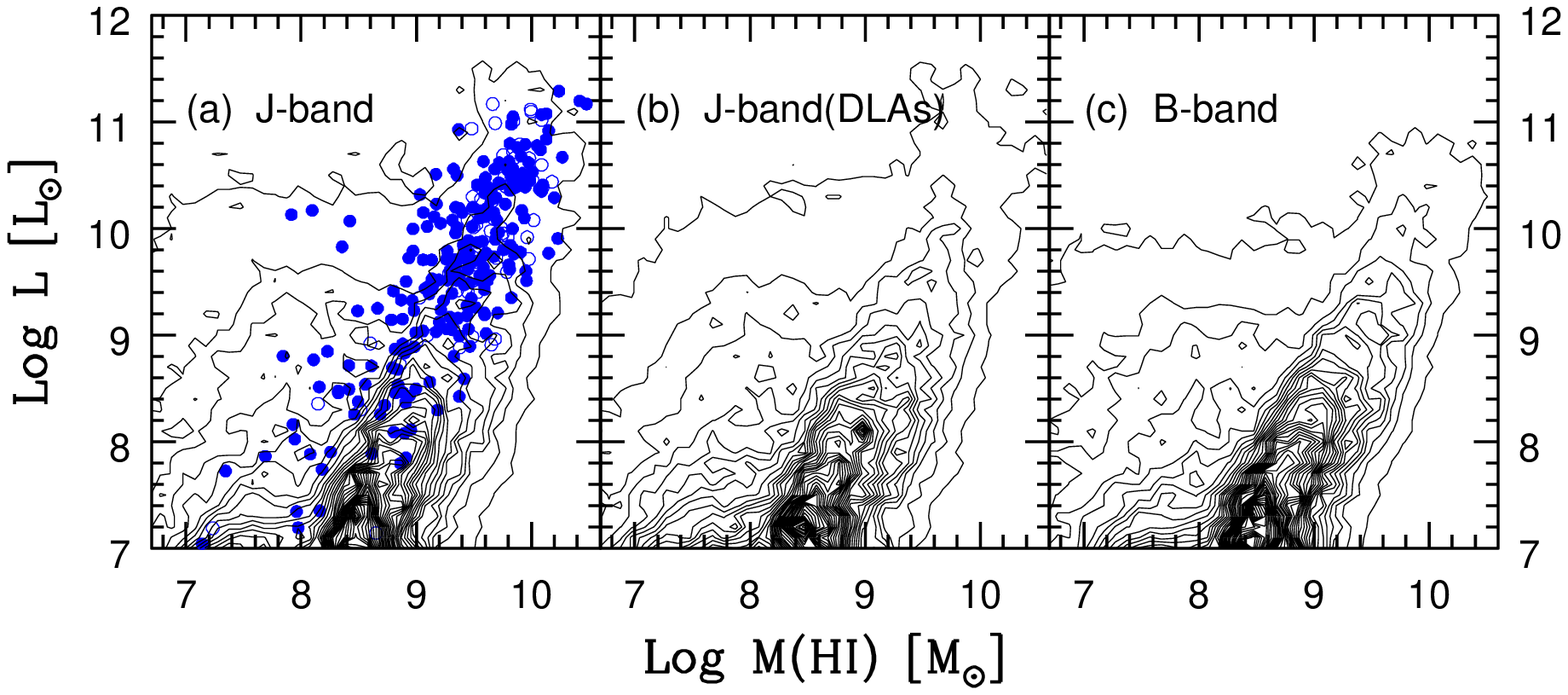}

\caption{
Luminosities of galaxies and DLA galaxies as a function of HI mass at redshift $z=0$ for the LS model. 
The contour levels are shown in order from outer to inner at an interval of $10^{-4}$ 
based on the calculated number of galaxies and DLA galaxies per logarithm of luminosity per logarithm 
of HI mass.  
(a) The contour map for the HI mass, $M_{\rm HI}$, vs the luminosity in the $J$ band, $L_{\rm J}$, 
of galaxies.  The symbols indicate the observational data from ADBS data (filled circles, Rosenberg \& Schneider 2000; open circles, Spitzak \& Schneider 1998). 
(b) The contour map for $M_{\rm HI}$ vs $L_{\rm J}$ of DLA galaxies. 
(c) The contour map for $M_{\rm HI}$ vs the luminosity in the $B$ band $L_{\rm B}$ of galaxies. 
} 
\label{fig:corrfil}
\end{figure*}

\begin{figure}
\plotone{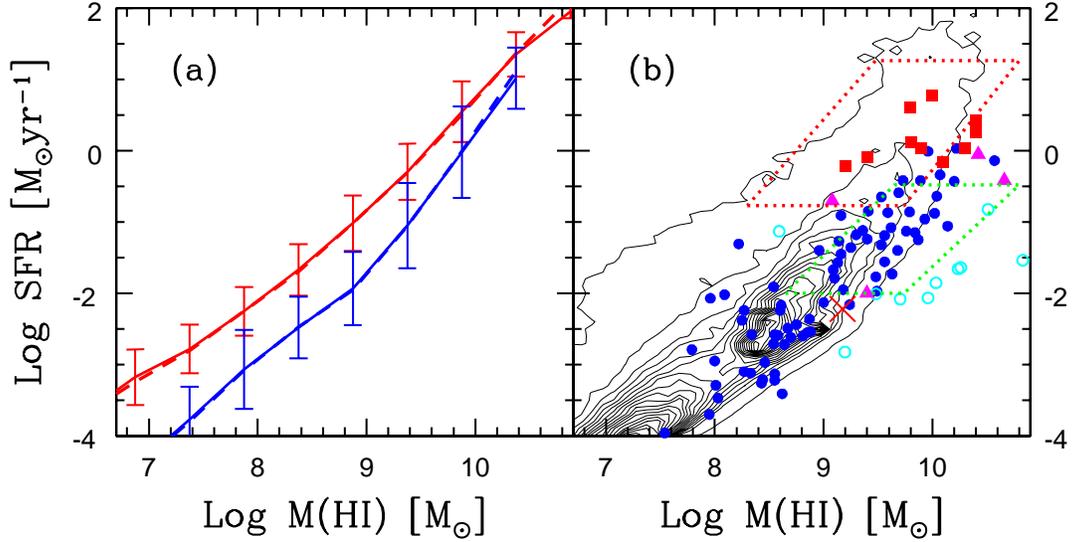}

\caption{ (a) Star formation rates in galaxies as a function of HI mass at redshift $z=0$ 
for the HS model (red solid line) and the LS model (blue solid line). 
Error bars with the averages indicate one $\sigma$ errors. The SFRs in DLA galaxies 
are also shown for the HS model (red dashed line) and the LS  model (blue dashed line).  
(b) The contour map for the SFR vs. the HI mass $M_{\rm HI}$ in 
galaxies for the LS model. 
The contour levels are shown in order from outer to inner at an interval of $3 \times 10^{-4}$ 
based on the calculated number of galaxies per logarithm of SFR per logarithm of HI mass.  
The star formation rates of the sample from the Survey for Ionization in Neutral Gas Galaxies 
are shown as {\it filled circles}(Meurer et al. 2006). The SFRs from optical and infrared 
surveys are also presented as {\it open circles} (van Zee et al. 1997), {\it filled triangles} 
(O'Neil, Oey, \& Bothun. 2007) and {\it filled squares} (Rahman et al. 2007). The SFRs in the HOPCAT  galaxies obtained from the infrared survey (red dotted box) and the 
radio ones  (green dotted box) are also presented (Doyle \& Drinkwater 2006).  
A SFR in an optical counterpart of a DLA system is plotted by cross  
(Schulte-Ladbeck et al. 2004).   
}
\label{fig:corrfil}

\end{figure}

\begin{figure}
\plotone{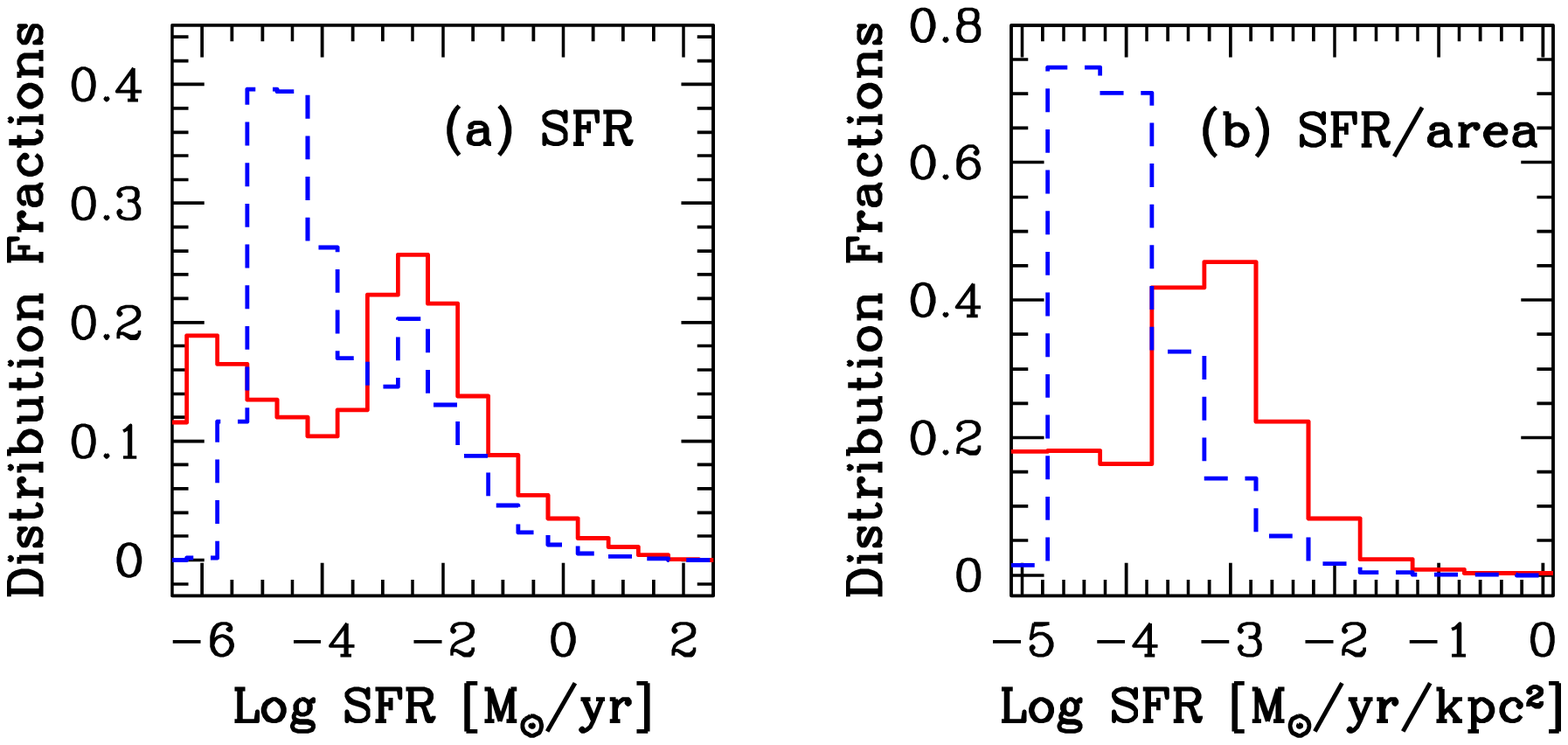}

\caption{ (a) Distribution functions of galaxies as a function of the logarithmic star 
formation rate 
($M_{\odot}$ yr$^{-1}$) for the HS model ({\it solid line}) and 
the LS model ({\it dashed line}). 
(b) The distribution functions of galaxies as a function of the logarithmic star formation 
rate per unit area ($M_{\odot}$ yr$^{-1}$ kpc$^{-2}$]) for the HS model ({\it solid line}) and 
the LS model ({\it dashed line}). 
}   
\label{fig:corrfil}

\end{figure}

\begin{figure}
\plotone{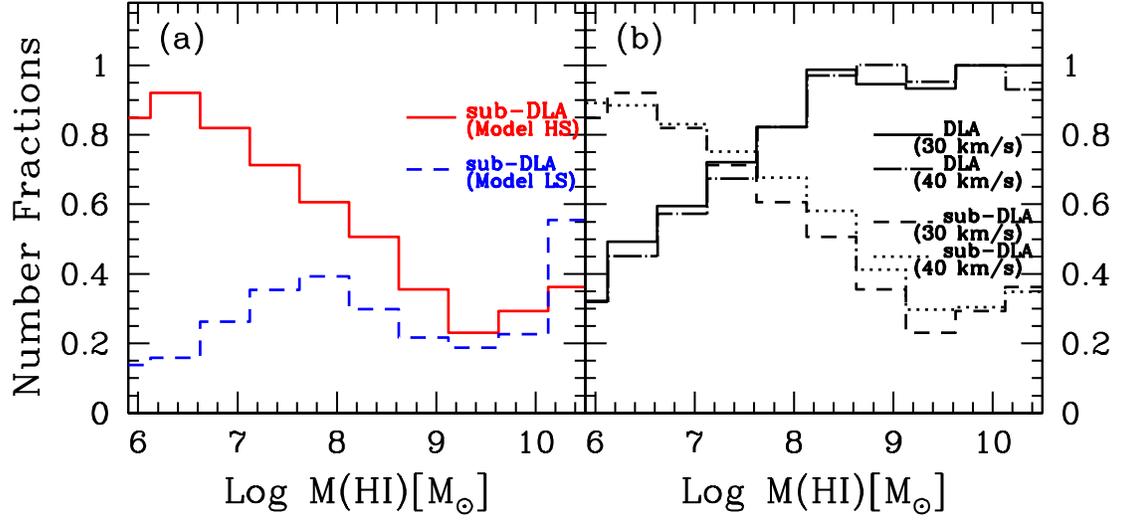}

\caption{(a) Histogram of number fractions of sub-DLA galaxies for the HS model 
({\it solid line}) and the LS model ({\it dashed line}). 
The number fraction at each bin of HI mass is defined as a ratio of the number density of 
sub-DLA galaxies relative to that of galaxies per bin of HI mass. 
(b) The histogram for number fractions of DLA and sub-DLA galaxies with the different 
thresholds for the halo circular velocity, V$_{\rm circ,th}$,  in the HS model. 
In the case of  V$_{\rm circ,th}=30$ km s$^{-1}$, 
the number fractions of DLA and sub-DLA galaxies are shown as {\it solid line} and 
dashed lines, respectively. In the case of V$_{\rm circ,th}=40$ km s$^{-1}$, 
the number fractions of DLA and sub-DLA galaxies are shown as dash-dotted  
dotted lines, respectively. 
} 
\label{fig:corrfil}
\end{figure}

\begin{figure}
\plotone{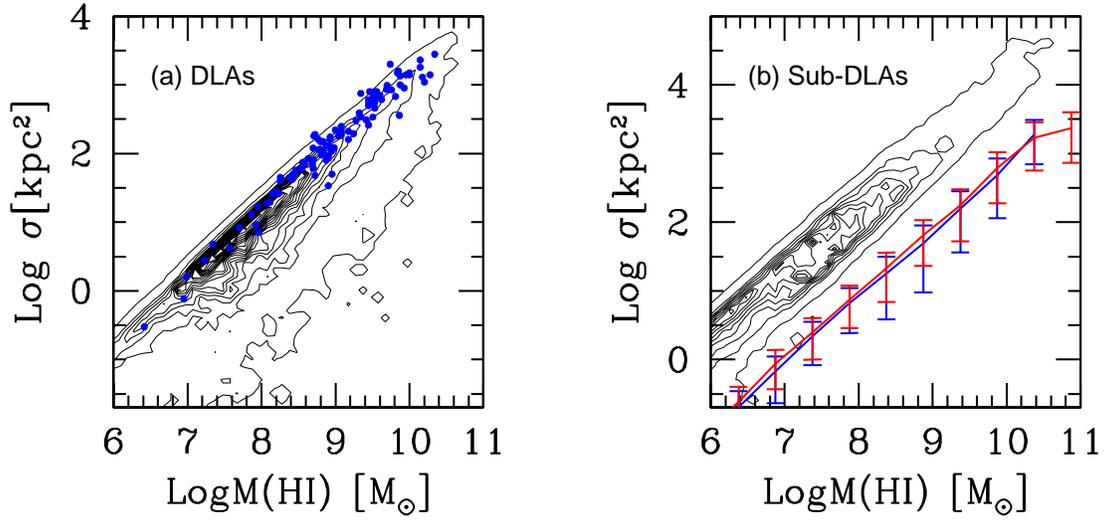}

\caption{ 
(a) Contour map for HI mass $M_{\rm HI}$ vs the cross section 
$\sigma$ of DLA galaxies for the HS model. 
The contour levels are shown in order from outer to inner at an interval of $2 \times 10^{-4}$ 
based on the calculated number of galaxies per logarithm of cross section per logarithm of HI mass.  
The dots represent the observational data provided by a blind radio survey 
(Rosenberg \& Schneider 2003). 
(b) The contour map for $M_{\rm HI}$ vs. $\sigma$ of sub-DLA galaxies 
for the HS model. The contour levels are the same as those presented in panel (a). 
For comparison, the mean cross sections of DLA galaxies with $1 \sigma$ error bars are also 
 shown for the HS model  (red line) and the LS model (blue line). 
} 
\label{fig:corrfil}
\end{figure}

\begin{figure}
\plotone{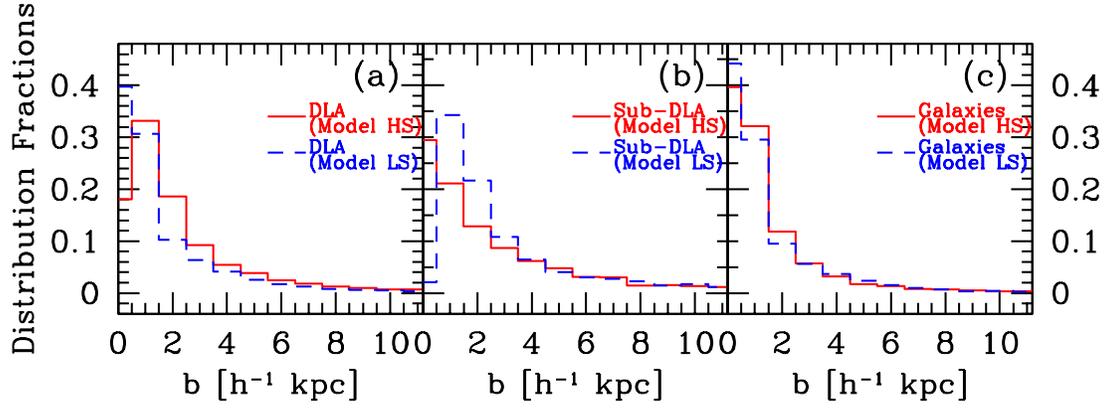}

\caption{(a) Distribution functions of (a) DLA galaxies, 
(b) sub-DLA galaxies, and (c) galaxies  as a function of the radii  
for the HS model ({\it solid line}) and the LS model 
({\it dashed line}). 
The number fraction is defined as a ratio of the galactic number 
within each bin of a radius $b$ to the total number.  
}

\end{figure}

\begin{figure}
\plotone{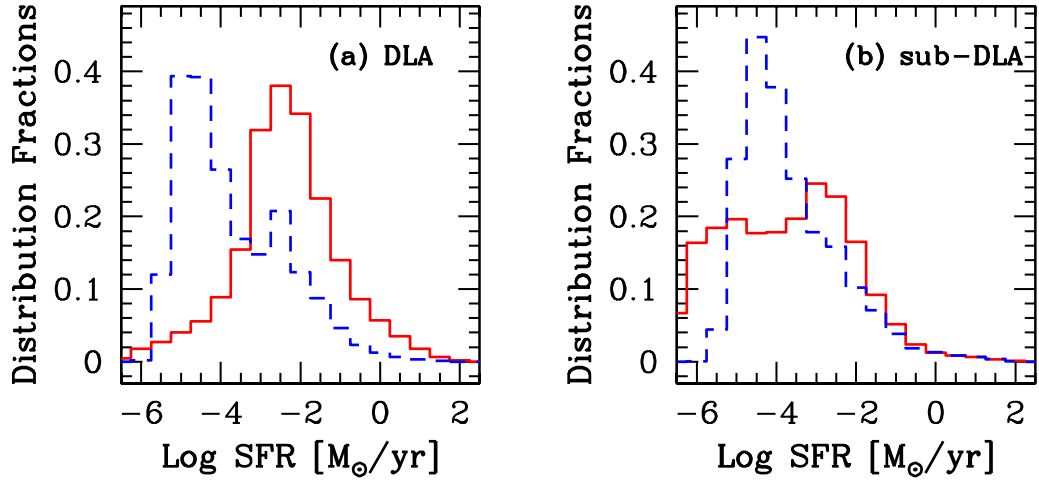}

\caption{(a) Distribution functions of DLA galaxies as a function of the logarithmic star 
formation rate ($M_{\odot}$ yr$^{-1}$]) for the HS model (solid line) 
and the LS model ({\it dashed line}). 
(b)The distribution functions of sub-DLA galaxies as a function of the logarithmic 
star formation rate ($M_{\odot}$ yr$^{-1}$) for the HS model (solid line) 
and the LS model (dashed line).  
}   
\end{figure}

\begin{figure}
\plotone{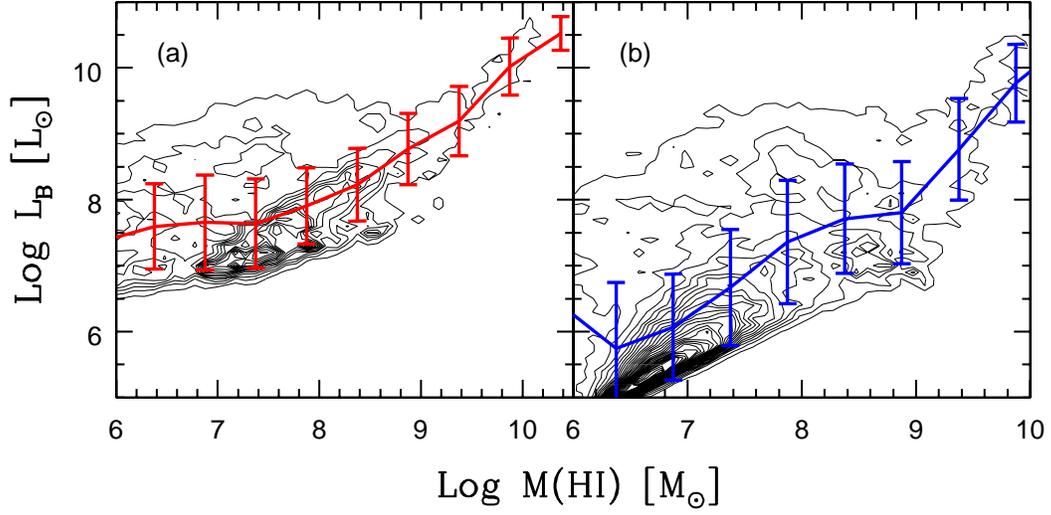}

\caption{Contour map for $M_{\rm HI}$ vs $L_{\rm B}$ of sub-DLA
 galaxies for (a) the HS model and (b) the LS model, respectively. 
The contour levels are shown in order from outer to inner at an interval of $10^{-4}$ 
based on the calculated number of galaxies and DLA galaxies per logarithm of the 
$B$-band luminosity per logarithm of HI mass. 
The mean luminosities with $1 \sigma$ error bars are shown as solid  lines. 
} 
\end{figure}

\end{document}